\documentclass[apj]{emulateapj}

\submitted{ApJ, accepted 07/08/2009}

\usepackage{amsmath}
\usepackage{import}
\usepackage{multirow}

\newcommand{\be}{\begin{equation}}
\newcommand{\ee}{\end{equation}}
\newcommand{\bea}{\begin{eqnarray}}
\newcommand{\eea}{\end{eqnarray}}
\newcommand{\bes}{\be\begin{split}}
\newcommand{\eesp}{\end{split}\ee}
\newcommand{\ha}{HI}
\newcommand{\hm}{H$_2$}

\newcommand{\abs}[1]{|#1|}

\newcommand{\h}{h}
\newcommand{\freq}{\nu}

\newcommand{\fco}{\freq_{\rm CO}}

\newcommand{\qha}{q_{\rm HI}}
\newcommand{\qhm}{q_{\rm H_2}}
\newcommand{\qnha}{q_{\rm HI,0}}
\newcommand{\qnhm}{q_{\rm H_2,0}}

\newcommand{\hp}{h_{\rm p}}

\newcommand{\f}{R_{\rm mol}}
\newcommand{\fc}{\f^{\rm c}}

\newcommand{\rd}{r_{\rm disk}}

\newcommand{\vg}{\sigma}

\newcommand{\mass}{M}
\newcommand{\msun}{{\rm M}_{\odot}}

\newcommand{\mha}{\mass_{\rm HI}}
\newcommand{\mhm}{\mass_{{\rm H}_2}}

\newcommand{\Sigmaha}{\Sigma_{\rm HI}}
\newcommand{\Sigmahm}{\Sigma_{\rm H_2}}

\newcommand{\Sigmastd}{\tilde{\Sigma}_{\rm H}}
\newcommand{\Omegaha}{\Omega_{\rm HI}}

\newcommand{\lha}{L_{\rm HI}}

\newcommand{\lsun}{{\rm L}_{\odot}}

\newcommand{\zc}{z_{\rm c}}

\newcommand{\dc}{D_{\rm C}}
\newcommand{\dl}{D_{\rm L}}
\newcommand{\da}{D_{\rm A}}
\newcommand{\dcmax}{D_{\rm C,max}}
\newcommand{\zmax}{z_{\rm max}}
\newcommand{\lfreq}{L}

\newcommand{\sfreq}{S}
\newcommand{\svel}{S^{\rm V}}

\newcommand{\rx}{r_{\rm x}}
\newcommand{\ry}{r_{\rm y}}
\newcommand{\rz}{r_{\rm z}}
\newcommand{\dndz}{{\rm d}N/{\rm d}z}
\newcommand{\simbox}{s_{\rm box}}
\newcommand{\wobs}[1]{{w_{\rm #1}^{\rm obs}}}

\shorttitle{HI and CO Observing Cones}
\shortauthors{Obreschkow et al.}

\begin{document}

\title{A Virtual Sky with Extragalactic HI and CO Lines for the SKA and ALMA}

\author{D. Obreschkow$^1$, H.-R. Kl\"ockner$^1$, I. Heywood$^1$, F. Levrier$^2$, and S. Rawlings$^1$}
\affil{$^1$\,Astrophysics, Department of Physics, University of Oxford, Keble Road, Oxford, OX1 3RH, UK\\$^2$\,Ecole Normale Sup\'{e}rieure, LERMA-LRA, UMR 8112 du CNRS, 24 rue Lhomond 75231, Paris Cedex 05}

\begin{abstract}
We present a sky simulation\footnote{http://s-cubed.physics.ox.ac.uk/, go to ``S$^3$-SAX-Sky''} of the atomic \ha~emission line and the first ten $\rm ^{12}C^{16}O$ rotational emission lines of molecular gas in galaxies beyond the Milky Way. The simulated sky field has a comoving diameter of $500\,\h^{-1}\rm\,Mpc$, hence the actual field-of-view depends on the (user-defined) maximal redshift $\zmax$; e.g.~for $\zmax=10$, the field of view yields $\sim4\times4\rm\,deg^2$. For all galaxies, we estimate the line fluxes, line profiles, and angular sizes of the \ha~and CO emission lines. The galaxy sample is complete for galaxies with cold hydrogen masses above $10^8\,\msun$. This sky simulation builds on a semi-analytic model of the cosmic evolution of galaxies in a $\Lambda$-cold dark matter ($\Lambda$CDM) cosmology. The evolving CDM-distribution was adopted from the Millennium Simulation, an $N$-body CDM-simulation in a cubic box with a side length of $500\,\h^{-1}\rm\,Mpc$. This side length limits the coherence scale of our sky simulation: it is long enough to allow the extraction of the baryon acoustic oscillations (BAOs) in the galaxy power spectrum, yet the position and amplitude of the first acoustic peak will be imperfectly defined. This sky simulation is a tangible aid to the design and operation of future telescopes, such the SKA, the LMT, and ALMA. The results presented in this paper have been restricted to a graphical representation of the simulated sky and fundamental $\dndz$-analyzes for peak flux density limited and total flux limited surveys of \ha~and CO. A key prediction is that \ha~will be harder to detect at redshifts $z\gtrsim2$ than predicted by a no-evolution model. The future verification or falsification of this prediction will allow us to qualify the semi-analytic models.
\end{abstract}


\section{Introduction}\label{section_introduction}

The interstellar medium (ISM) is the bridge between the environment of galaxies and their newborn stars. Its atomic and molecular phases can be detected via emission lines. Typically studied lines include the \ha-radio line ($1.420\rm~GHz$ rest-frame) and the rotational CO-lines in the (sub)millimeter spectrum (multiples of $115.27\rm~GHz$). These lines characterize both the composition and the dynamical state of the ISM, and their apparent frequency measures the redshift of the source. If the object sits at a cosmological distance, the redshift is dominated by the expansion of the Universe and hence provides a distance measure. Therefore, observations of \ha~and CO at high redshift are currently discussed as a means of localizing high-redshift galaxies \citep{Carilli2002b,Carilli2004b}, thus unveiling an unprecedented image of cosmic structure \citep[e.g.][]{Abdalla2008}.

However, due to current sensitivity limitations, no \ha-emission has yet been found beyond redshift $z=0.25$ \citep{Verheijen2007,Catinella2008}. In contrast, CO-emission lines have been detected in different systems out to $z\approx6.4$ \citep{Walter2004}, yet all these lines originate from atypical objects, such as ultra luminous infrared galaxies (ULIRGs) or quasi stellar objects (QSOs), with the exception of two ordinary galaxies recently detected in CO(2--1)-emission at $z\approx1.5$ \citep{Daddi2008}. Both \ha~\citep{Prochaska2005} and \hm~\citep{Noterdaeme2008} have also been measured via absorption of their respective Lyman lines against distant QSOs. Yet, the nature of the absorbing galaxies remains unknown.

The discrepancy between the primordial astrophysical importance of cold gas in galaxies and its gravely limited detectability at high redshift is a main driver for the design of many future radio and (sub)millimeter telescopes. Prominent examples are the Square Kilometre Array (SKA), the Large Millimeter Telescope (LMT), and the Atacama Large Millimeter/submillimeter Array (ALMA), which are expected to detect \ha~and CO at high $z$. The optimization of these instruments and the planning of their surveys require robust predictions of the detectable signatures. Such predictions are available for the continuum radiation of a large sample of galaxies \citep[e.g.][]{Wilman2008}. By contrast, high-redshift line emission of \ha~and CO has only been simulated for single galaxies \citep{Boomsma2002,Combes1999,Greve2008} or simplistically extrapolated for a population of regular galaxies \citep{Blain2000,Carilli2002b,Abdalla2005}. All these models ignore the implications of galaxy mergers, cooling flow suppression mechanisms, and other complex phenomena. Moreover, most line simulations exclusively consider either the atomic or the molecular gas phase of the ISM. This approach implicitly assumes that the other phase is negligible or that the masses in both phases evolve proportionally. Both of these assumptions contradict recent studies of the co-evolution of \ha~and \hm~in regular galaxies \citep{Obreschkow2009c}. Finally, the line simulations cited above neglect cosmic large-scale structure. The time seems ripe for joint predictions of atomic and molecular emission lines in a sample of galaxies large enough to probe cosmic structure.

In this paper, we present a simulation of a sky field with a comoving diameter of $500\,\h^{-1}\rm\,Mpc$. The actual field-of-view depends on the (user-defined) maximal redshift $\zmax$; e.g.~for $\zmax=1$ the field of view yields $\sim12\times12\rm\,deg^2$, or for $\zmax=10$ the field of view yields $\sim4\times4\rm\,deg^2$. This simulation is obtained by constructing a mock observing cone from a previously presented galaxy simulation. The latter relies on the large-scale structure computed by the Millennium Simulation \citep{Springel2005} and an enhanced semi-analytic galaxy model \citep{Croton2006,DeLucia2007,Obreschkow2009b}.

Section \ref{section_methods} explains the simulation methods. In Section \ref{section_results}, we provide a graphical illustration of the simulated sky field and extract $\dndz$-estimates for peak flux density limited surveys. Section \ref{section_discussion} discusses some important limitations of the presented simulation. A non-exhaustive list of possible applications is provided in Section \ref{section_conclusion} along with a brief conclusion. The appendices show additional illustrations, list the parameters describing the analytic fits to the predicted $\dndz$-functions, and describe the on-line access to the simulation data.

\section{Methods}\label{section_methods}

In this section, we describe the multiple simulation steps required to progress from a simulation of the \emph{evolution} of cosmic structure to a \emph{static} sky simulation. We have grouped this description into four steps, corresponding to four successive simulation steps. The first step (Section \ref{subsection_galaxysim}) contains all the simulation work presented in earlier studies. This work resulted in a catalog of $\sim3\cdot10^7$ evolving galaxies with detailed cold gas properties. In the second step (Section \ref{subsection_mockcone}), this catalog is transformed into a mock observing cone, which represents a virtual sky field. In the third step (Section \ref{subsection_linefluxes}), the intrinsic properties of the galaxies in this virtual sky field are converted into apparent line fluxes. In the fourth step (Sections \ref{subsection_line_profiles} and \ref{subsection_angular_sizes}), the line emission is refined by the evaluation of line profiles and angular sizes of the line-emitting gas.

\subsection{Simulation of the ISM in $\sim\!3\!\cdot\!10^7$ evolving galaxies}\label{subsection_galaxysim}

Here, we recapitulate the galaxy simulation presented in earlier studies. This simulation relies on three consecutive layers: (i) a simulation of the cosmic evolution of dark matter \citep{Springel2005}; (ii) a semi-analytic simulation of the evolution of galaxies on the dark matter skeleton \citep{Croton2006,DeLucia2007}; and (iii) a post-processing to split the cold hydrogen masses associated with each galaxy into \ha~and \hm~\citep{Obreschkow2009b}.

For the dark matter simulation, we adopted the Millennium Simulation \citep{Springel2005}, an $N$-body dark matter simulation within the standard $\Lambda$-cold dark matter ($\Lambda$CDM) cosmology. This simulation uses a cubic simulation box with periodic boundary conditions and a comoving volume of $(500\,\h^{-1}\,{\rm Mpc})^3$. The Hubble constant was fixed to $H_0=100\,\h\rm\,km\,s^{-1}\,Mpc^{-1}$ with $\h=0.73$. The other cosmological parameters were chosen as $\Omega_{\rm matter}=0.25$, $\Omega_{\rm baryon}=0.045$, $\Omega_\Lambda=0.75$, and $\sigma_8=0.9$. The simulation-box contains $\sim10^{10}$ particles with individual masses of $8.6\cdot10^8\,\msun$. This mass resolution allows the identification of structures as low in mass as the Small Magellanic Cloud \citep[see][]{Springel2005}.

For the second simulation-layer, i.e.~the cosmic evolution of the galaxies distributed on the dark matter skeleton, we adopt the semi-analytic model of \cite{DeLucia2007} \citep[see~ also][]{Croton2006}. In this macroscopic model all galaxies are represented by a list of global properties, such as position, velocity, and total masses of gas, stars, and black holes. These properties are evolved using empirically or theoretically motivated formulae for mechanisms, such as gas cooling, reionization, star formation, gas heating by supernovae, starbursts, black hole accretion, black hole coalescence, and the formation of stellar bulges via disk instabilities. The resulting virtual galaxy catalog (hereafter the ``DeLucia-catalog'') contains the positions, velocities, merger histories, and intrinsic properties of $\sim3\cdot10^7$ galaxies at 64 cosmic time steps. The free parameters in the semi-analytic model were tuned to various observations in the local universe (see \citealp{Croton2006}). Therefore, despite the simplistic implementation and the possible incompleteness of this model, the DeLucia-catalog nonetheless provides a good fit to the joint luminosity/colour/morphology distribution of observed low-redshift galaxies \citep{Cole2001,Huang2003,Norberg2002}, the bulge-to-black hole mass relation \citep{Haering2004}, the Tully--Fisher relation \citep{Giovanelli1997}, and the cold gas metallicity as a function of stellar mass \citep{Tremonti2004}.

In this paper, we are particularly interested in the cold gas masses of the galaxies in the DeLucia-catalog. These cold gas masses are the net result of (i) gas accretion by cooling from a hot halo (dominant mode) and galaxy mergers, (ii) gas losses by star formation and feedback from supernovae, (iii) and cooling flow suppression by feedback from accreting black holes. The DeLucia-catalog does not distinguish between molecular and atomic cold gas, but simplistically treats all cold gas as a single phase. The atomic and molecular phases are therefore dealt with in the third simulation layer.

The third simulation-layer, i.e.~the subdivision of the cold hydrogen mass of each galaxy into \ha- and \hm-distributions \citep{Obreschkow2009b}, relies on an analytic model for the mass-distributions of \ha~and \hm~within regular galaxies. In this model, the column densities of \ha~and \hm, $\Sigmaha$ and $\Sigmahm$ respectively, are given by
\bea
  \Sigmaha(r) & = & \frac{\Sigmastd\,\exp(-r/\rd)}{1+\fc\exp(-1.6\,r/\rd)}\,, \label{eqsigmaha} \\
  \Sigmahm(r) & = & \frac{\Sigmastd\,\fc\,\exp(-2.6\,r/\rd)}{1+\fc\exp(-1.6\,r/\rd)}\,, \label{eqsigmahm}
\eea
where $r$ denotes the galactocentric radius, $\rd$ is a scale length, $\fc$ is the \hm/\ha-mass ratio at the galaxy center, and $\Sigmastd$ is a normalization factor. We derived Eqs.~(\ref{eqsigmaha},\ref{eqsigmahm}) based on a list of empirically supported assumptions, the most important of which are: (i) the cold gas of regular galaxies resides in a flat disk (see \citealp{Leroy2008} for local spiral galaxies, \citealp{Young2002} for local elliptical galaxies, \citealp{Tacconi2006} for galaxies at higher redshifts); (ii) the surface density of the total hydrogen component (\ha+\hm) is well described by an axially symmetric exponential profile \citep{Leroy2008}; (iii) the local \hm/\ha-mass ratio scales as a power of the gas pressure of the ISM outside molecular clouds \citep{Blitz2006}.

Using Eqs.~(\ref{eqsigmaha},\ref{eqsigmahm}), we can characterize the \ha- and \hm-content of every simulated galaxy in the DeLucia-catalog. The resulting hydrogen simulation successfully reproduces many local observations of \ha~and \hm, such as mass functions (MFs), mass--diameter relations, and mass--velocity relations \citep{Obreschkow2009b}. This success is quite surprising, since our model for \ha~and \hm~only introduced one additional free parameter to match the observed average space density of cold gas in the local Universe \citep{Obreschkow2009b}. A key prediction of this simulation is that the \hm/\ha-ratio of most regular galaxies increases dramatically with redshift, hence causing a clear signature of cosmic downsizing in the \hm-MF \citep{Obreschkow2009c}.

Despite its consistency with existing observations, we emphasize that the presented model for the cosmic evolution of \ha~and \hm~is simplistic and uncertain. In particular at high $z$, the model assumptions may significantly differ from the reality. For example, high-$z$ galaxies are likely to be more disturbed due to higher merger rates and long dynamic time scales compared to their age. There is also evidence that cold gas disks become more turbulent with redshift \citep[e.g.][]{ForsterSchreiber2006,Genzel2008}. Uncertainties from these and other model limitations are discussed briefly in Section \ref{section_discussion} and in depth in Section 6 of \citet{Obreschkow2009b}.

\subsection{Building a mock observing cone}\label{subsection_mockcone}

We shall now describe how the cubic simulation-box is transformed into a virtual sky field. This procedure can be regarded as a fourth simulation-layer on top of the hierarchical simulation described in Section \ref{subsection_galaxysim}.

\begin{figure}[h]
  \includegraphics[width=\columnwidth]{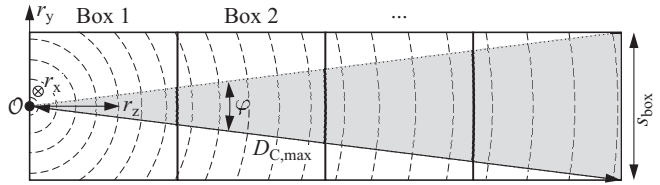}
  \caption{Schematic illustration of the construction of a mock observing cone (shaded region) from a chain of replicated simulation boxes (solid square boxes). The galaxies in the replicated boxes are drawn from the discrete cosmic-time step of the simulation, which best corresponds to their cosmological redshift relative to the observer $\mathcal{O}$. Galaxies from the same discrete time step therefore lie within spherical shells around the observer (indicated by the dashed-lines).}
  \label{fig_mockmap_technique}
\end{figure}

The method adopted here closely follows the one described by \cite{Blaizot2005}, namely the building of a chain of replicated simulation-boxes along the line-of-sight, as shown in Fig.~\ref{fig_mockmap_technique}. At any position in this chain, the galaxies are drawn from the cosmic time in the simulation, which corresponds to the look-back time, at which the galaxy is seen by the observer $\mathcal{O}$. Since our galaxy simulation uses 64 discrete time steps, we describe each galaxy in the cone by its properties at the closest available time step\footnote{The galaxy properties cannot readily be interpolated between two successive time steps, since a galaxy at any time step may have several progenitors.}, in terms of redshift. This defines the spherical shells of identical cosmic time, which are separated by dashed lines in Fig.~\ref{fig_mockmap_technique}. The relatively narrow redshift-separation of these shells ensures that the assigned galaxy properties cannot differ significantly from the properties at their exact look-back time.

The same galaxy can appear once in every box in Fig.~\ref{fig_mockmap_technique}, but with different intrinsic properties due to the cosmic evolution. However, the position of the repeated galaxy in comoving coordinates will be very similar, which can result in spurious radial features for the observer $\mathcal{O}$ \citep[see Fig.~1 in][]{Blaizot2005}. To suppress this effect, we randomize the galaxy positions by applying random symmetry operations to each box in the chain. These operations consist of 90${\rm\,deg}$-rotations, inversions, and continuous translations\footnote{Translations can be applied because of the periodic boundary conditions imposed on the simulation box of the Millennium Simulation.}. Applying these symmetry operations also removes the non-physical periodicity of $500\,\h^{-1}\,{\rm Mpc}$ associated with the side length of the periodic simulation-box. But we emphasize that applying the symmetry operations does not provide information on scales larger than the simulation-box. Symmetry operations can, however, introduce unwanted small-scale density variations at the interface of two neighboring boxes. These and other limitations of this method are discussed by \cite{Blaizot2005}.

From the randomized chain of replicated simulation-boxes, an observing cone can be extracted (shaded region in Fig.~\ref{fig_mockmap_technique}). Each galaxy in this cone is projected onto the celestial sphere centered about the vernal point (${\rm RA}=0$, ${\rm Dec}=0$). The Euclidian projection formulas for arbitrary large angles are

\bea
  {\rm RA}  & = & \arctan\left(\frac{\rx}{\rz}\right), \label{eqra} \\
  {\rm DEC} & = & \arctan\left(\frac{\ry}{\sqrt{\rx^2+\rz^2}}\right) \label{eqdec},
\eea
where $\rx$, $\ry$, and $\rz$ are the comoving coordinates of the galaxy relative to the observer (see Fig.~\ref{fig_mockmap_technique}). The ``cosmological redshift'' $z$ of each galaxy is computed directly from its comoving distance $\dc=(\rx^2+\ry^2+\rz^2)^{1/2}$, while the Doppler-shift corrected ``apparent redshift'' is computed as $\tilde{z}=z+V_{\rm r}/c$, where $V_{\rm r}$ is the peculiar recession velocity of the galaxy relative to the Hubble flow.

Fig.~\ref{fig_mockmap_technique} shows that the opening angle $\varphi$ of the virtual sky field is set by the maximal comoving distance $\dcmax$ via
\be\label{eqphiboxes}
  \varphi = 2\,\arcsin\frac{\simbox}{2\,\dcmax},
\ee
where $\simbox$ is the comoving side length of the simulation-box. Given a value of $\simbox$ and a choice of cosmological parameters, Eq.~(\ref{eqphiboxes}) implies a one-to-one relation between $\varphi$ and the maximal redshift $z_{\rm max}$.

\begin{figure}[h]
  \includegraphics[width=\columnwidth]{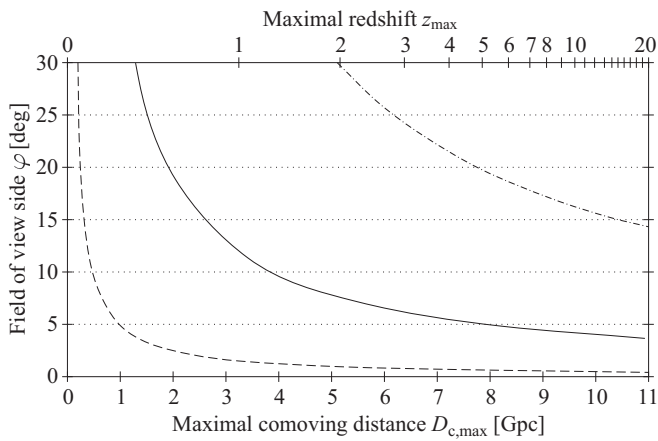}
  \caption{Relation between the maximal comoving distance $\dcmax$ or maximal redshift $\zmax$ and the opening angle $\varphi$ of the mock observing cone. The different lines correspond to the box sizes of the Millennium Simulation (solid), the Milli-Millennium Simulation (dashed), and the Horizon-$4\pi$ Simulation (dash-dotted).}
  \label{fig_fov}
\end{figure}

Fig.~\ref{fig_fov} displays the relation between $\varphi$, $\dcmax$, and $\zmax$ for the cosmological parameters of the Millennium Simulation (Section \ref{subsection_galaxysim}) and three different side lengths $\simbox$. The choice $\simbox=500\,\h^{-1}\,{\rm Mpc}$ (solid line) corresponds to the box of the Millennium Simulation. $\simbox=62.5\,\h^{-1}\,{\rm Mpc}$ (dashed line) corresponds to the ``Milli-Millennium'' Simulation, a small test version of the Millennium Simulation. $\simbox=2\,\h^{-1}\,{\rm Gpc}$ (dash-dotted line) corresponds to the giant simulation-box of the Horizon-$4\pi$ Simulation, a dark matter stimulation with $\sim10$-times less mass resolution than the Millennium Simulation (\citealp{Prunet2008,Teyssier2008}; see also Section \ref{subsection_structure}).

\subsection{Assigning apparent line fluxes}\label{subsection_linefluxes}

We shall now assign apparent line fluxes to each galaxy in the mock observing cone constructed in Section \ref{subsection_mockcone}. In general, the frequency-integrated line flux $\sfreq$ of any emission line, can be computed from the frequency-integrated luminosity (= intrinsic power) via
\be\label{eqlftosf}
  \sfreq = \frac{\lfreq}{4\pi\dl^2},
\ee
where $\dl=(1+z)\,\dc$ is the luminosity distance to the source. We note that Eq.~(\ref{eqlftosf}) takes a different form for velocity-integrated fluxes and brightness temperatures. For reference, a summary of all relations between frequency-integrated and velocity-integrated fluxes, luminosities, and brightness temperatures has been compiled in Appendix A of \cite{Obreschkow2009e}.

In Sections \ref{subsubsection_haflux} and \ref{subsubsection_cofluxes}, we shall now summarize the models used to estimate the frequency-integrated luminosities $\lfreq$ (hereafter simply ``luminosities'') of different emission lines.

\subsubsection{Conversion of \ha-mass into \ha-line luminosities}\label{subsubsection_haflux}

We evaluate the \ha-luminosities $\lha$ from the \ha-masses $\mha$ of the simulated galaxies via the standard conversion \citep[e.g.][]{Meyer2004}
\be\label{eqlha}
  \frac{\lha}{\lsun}=6.27\cdot10^{-9}\cdot\frac{\mha}{\msun}.
\ee

The \ha-line or ``$21\rm\,cm$-line'' at a rest-frame frequency of $1.420\rm\,GHz$ originates from the photon-mediated transition between the two spin states of the proton-electron system in the electronic ground state. The upper spin state has a low spontaneous decay rate of $f=2.9\cdot10^{-15}\rm\,Hz$. This frequency is about 5 orders of magnitude smaller than that of \ha-\ha~collisions \citep[][]{Binney1998}. Hence, the two spin states are in thermal equilibrium with the kinetic state of the gas, which implies a spin-temperature far above the spin excitation temperature $T_{\rm ex}\approx0.07\,K$. Therefore, the spin systems are in the high-temperature limit, where $3/4$ of all systems are in the upper (three-fold degenerate) state. The radiative power emitted per atom can therefore be calculated as $\lha=0.75\,f\,\hp\,1.4{\rm\,GHz}$ ($\hp$ is the Planck constant), which readily reduces to Eq.~(\ref{eqlha}).

We have neglected \ha-self absorption, since this seems to affect only massive spiral galaxies when observed almost edge-on \citep{Ferriere2001,Wall2006}. This assumption should also be valid for high-redshift galaxies, since their \ha-masses were not much larger than today, as can be inferred from Lyman-$\alpha$ absorption measurements against distant quasars \citep{Lah2007} and as is predicted by our simulation \citep{Obreschkow2009c}.

\ha~in collapsed structures, i.e.~in galaxies, is generally warm enough ($\gtrsim50\rm\,K$) that the Cosmic Microwave Background (CMB) can be safely neglected as an observing background for all galaxies at $z<10$. Only in the intergalactic medium (IGM) during the cosmic Epoch of Reionization (EoR) can \ha~appear in $21\rm\,cm$-absorption against the CMB \citep[e.g.][]{Iliev2002}.

\subsubsection{\!\!Conversion of \hm-mass into CO-line luminosities}\label{subsubsection_cofluxes}

We derive the CO-line luminosities $L_{\rm CO}$ from the \hm-masses of the simulated galaxies. We only consider the radiation emitted by the most abundant CO-isotopomer, $\rm ^{12}C^{16}O$, when relaxing from one of the rotational states $J=1,...,10$ to the state $J-1$. The radiation frequency of such a decay is $J\cdot\fco$, where $\fco=115.27\rm\,GHz$ is the rest-frame frequency of the transition $J=1\!\rightarrow\!0$.

The conversion between \hm-masses and CO-luminosities is a highly nuanced affair with a long history in millimeter astronomy. We therefore presented an in-depth analysis of this conversion in \cite{Obreschkow2009e} and introduced a model to estimate the different luminosities $L_{\rm CO,J}$ of the galaxies in our simulation. This model respects the following physical mechanisms: (i) molecular gas is heated by the CMB, starbursts (SBs), and active galactic nuclei (AGNs); (ii) molecular clouds in dense or inclined galaxies can overlap; (iii) very dense gas is smooth instead of clumpy; (iv) the metallicity varies amongst galaxies and changes with redshift; (v) CO-luminosities are always detected against the CMB. We shall apply this model in the present paper. Limitations and uncertainties are discussed in Section \ref{subsection_limco}.

\subsection{Emission line profiles}\label{subsection_line_profiles}

Having assigned an integrated line flux to each galaxy in the simulation, we can now refine their attributes, by characterizing each line with a profile. To this end we depart from the \emph{edge-on} line profiles evaluated in \cite{Obreschkow2009b} for each galaxy in the simulation. We represented those profiles by normalized flux densities $\Psi(V)$, where $V$ is the velocity measured in the rest-frame of the center of the observed galaxy. The normalization condition, $\int_{\rm V}\Psi(V){\rm d}V=1$, implies that $\Psi(V)$ only needs to be multiplied by the velocity-integrated flux (in units of $\rm Jy\,km/s$) in order to obtain actual flux densities (in units of $\rm Jy$). For each galaxy we calculated two profiles $\Psi(V)$, one for the \ha-component and one for the \hm-component (associated with CO)\footnote{In \cite{Obreschkow2009b}, we called these two profiles $\Psi_{\rm HI}$ and $\Psi_{\rm CO}$, respectively.}. This calculation relied on a detailed mass model based on the halo, disk, and bulge of the galaxies, combined with our model for the \ha- and \hm-surface densities given in Eqs.~(\ref{eqsigmaha},\ref{eqsigmahm}). For practical reasons, the resulting line profiles were reduced to five parameters (see Fig.~\ref{fig_recover_profile}): the normalized flux density at the line center $\Psi_0$; the normalized peak flux density $\Psi_{\rm max}$, usually corresponding to the two peaks of a double-horn profile; the line width $w_{\rm peak}$ between the two peaks of the double-horn profile; the line width $w_{50}$ at the 50-percentile level of the peak flux density; and the line width $w_{20}$ at the 20-percentile level. The original normalized line profile can be approximately recovered from these parameters using the formulas in Appendix \ref{appendix_recovery}.

The remaining task consists of correcting the line profiles for the inclination $i$ of each galaxy\footnote{The simulated DeLucia-catalog does not provide galaxy orientations. We therefore assign inclinations randomly between $0\rm\,deg$ (face-on) and $90\rm\,deg$ (edge-on) according to a sine-distribution.}. $i$ is defined as the angle between the line-of-sight and the galaxy's rotation axis; hence $i=0{\rm\,deg}$ corresponds to face-on galaxies and $i=90{\rm\,deg}$ corresponds to edge-on galaxies. In the
absence of a random gas velocity dispersion, apparent line widths $w^{\rm obs}$
could be computed from the edge-on line widths $w$, via $w^{\rm obs}=w\cdot\sin{i}$, while apparent normalized flux densities would scale as $\Psi^{\rm obs}=\Psi/\sin{i}$.

Here, we assume that the cold gas has a random, isotropic velocity dispersion characterized by a Gaussian velocity distribution in each space dimension. The observed line profile of a face-on galaxy ($i=0{\rm~deg}$) then takes the shape of a Gaussian function. Under no inclination can the line profile become more narrow than this Gaussian function, nor can the normalized line flux densities become higher than the peak of this Gaussian function. Let $\vg$ be the standard-deviation of the Gaussian velocity dispersion. Then the minimum line widths are given by $w_{20}^{\rm obs}=2\sqrt{-2\,\ln(0.2)}\approx3.6\,\vg$, $w_{50}^{\rm
obs}=2\sqrt{-2\,\ln(0.5)}\approx2.4\,\vg$, and $w_{\rm peak}^{\rm obs}=0$, and the maximum normalized flux densities are
$\Psi_0=(\vg\sqrt{2\pi})^{-1}$ and $\Psi_{\rm max}=(\vg\sqrt{2\pi})^{-1}$.

In addition, the maximal normalized flux density $\Psi_{\rm max}$ cannot differ from the central normalized flux density $\Psi_0$ by an arbitrarily large amount, due to the line profile smoothing. Explicitly, the slope in the emission line between the points $\Psi_0$ and $\Psi_{\rm max}$ cannot exceed the maximal slope of the Gaussian velocity function, which is equal to $0.24\,\vg^{-2}$. This requirement translates into an upper bound for $\Psi_{\rm max}$ equal to $\Psi_{0}^{\rm obs}+0.12\,\vg^{-2}\,w_{\rm
peak}^{\rm obs}$. A set of equations respecting all of these conditions is given
by
\bea
  w_{20}^{\rm obs} & = & (w_{20}-3.6\,\vg)\cdot\sin{i}+3.6\,\vg, \label{eqline1} \\
  w_{50}^{\rm obs} & = & (w_{50}-2.4\,\vg)\cdot\sin{i}+2.4\,\vg, \label{eqline2} \\
  w_{\rm peak}^{\rm obs} & = & w_{\rm peak}\cdot\sin{i}, \label{eqline3} \\
  \Psi_{0}^{\rm obs} & = & \min\!\left(\!\frac{\Psi_{0}}{\sin{i}},\frac{1}{\vg\sqrt{2\pi}}\right), \label{eqline4} \\
  \Psi_{\rm max}^{\rm obs} & = & \min\!\left(\frac{\Psi_{\rm max}}{\sin{i}},\frac{1}{\vg\sqrt{2\pi}},\Psi_{0}^{\rm obs}\!+\!\frac{0.12w_{\rm peak}^{\rm obs}}{\vg^2}\!\right)\!. \label{eqline5}
\eea

For all the line profiles, we here adopt $\vg=8\rm\,km\,s^{-1}$ to remain consistent with \cite{Obreschkow2009b}. We note, however, that high-redshift galaxies may have higher velocity dispersions \citep{ForsterSchreiber2006} perhaps due to an intense ongoing accretion of gas. Another limitation of the presented line model is that all CO-lines have by definition the same line shape. This assumption nevertheless approximately agrees with simultaneous observations of different emission lines \cite[e.g.][]{Weiss2007,Greve2009}.

\begin{figure*}[t!]
  \includegraphics[width=\textwidth]{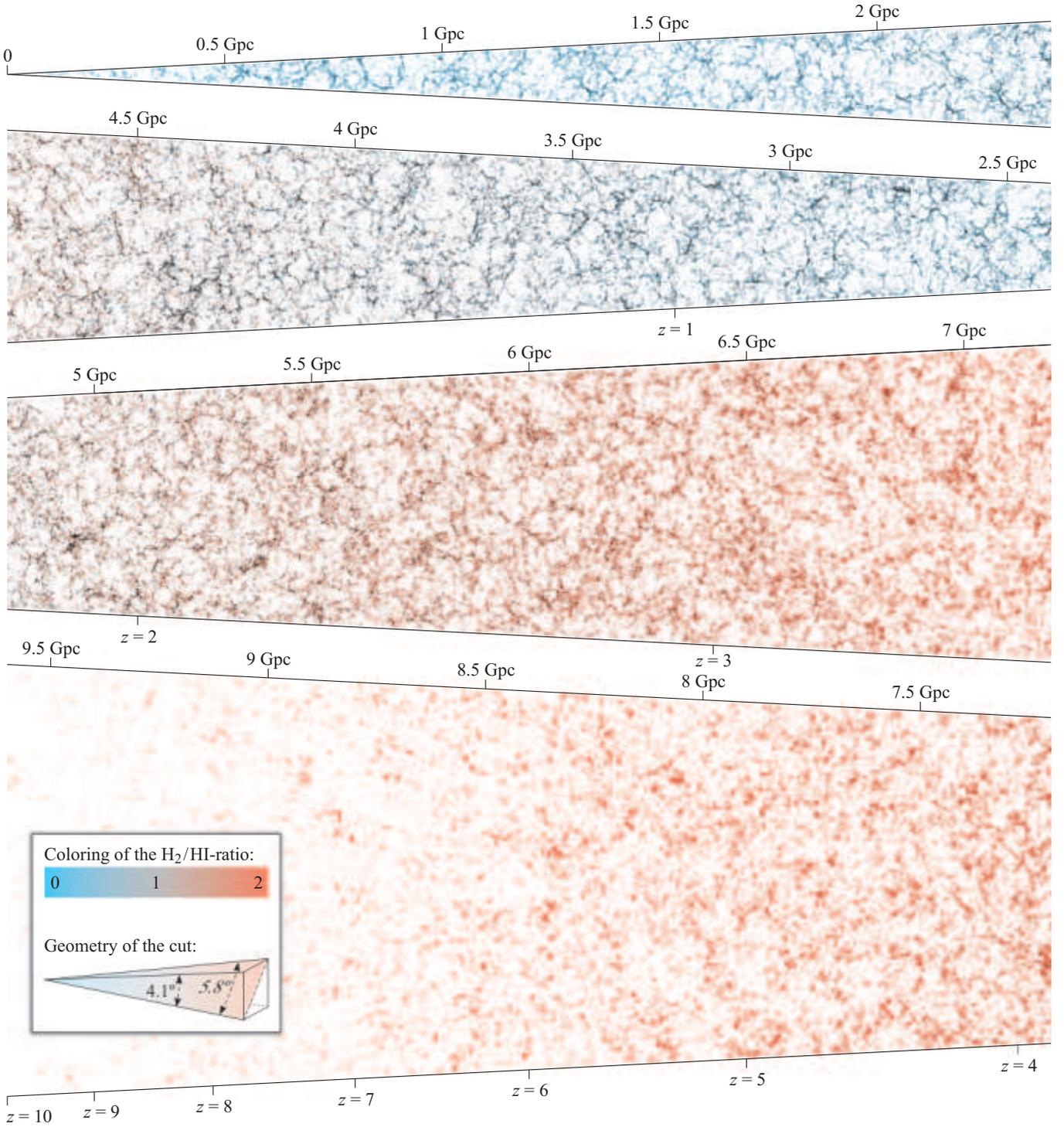}
  \caption{Longitudinal cut through the simulated mock observing cone. The cut slice has a thickness of $10\rm\,Mpc$ and is represented in comoving coordinates. For illustration purposes, the slice has been wrapped in four parts, which variably read from the left to the right and vice versa. The dots represent gas-rich galaxies and the coloring shows their \hm/\ha-ratio, from 0 (blue) to 2 (red). }\label{fig_cone}
\end{figure*}

\begin{figure*}[t!]
  \includegraphics[width=\textwidth]{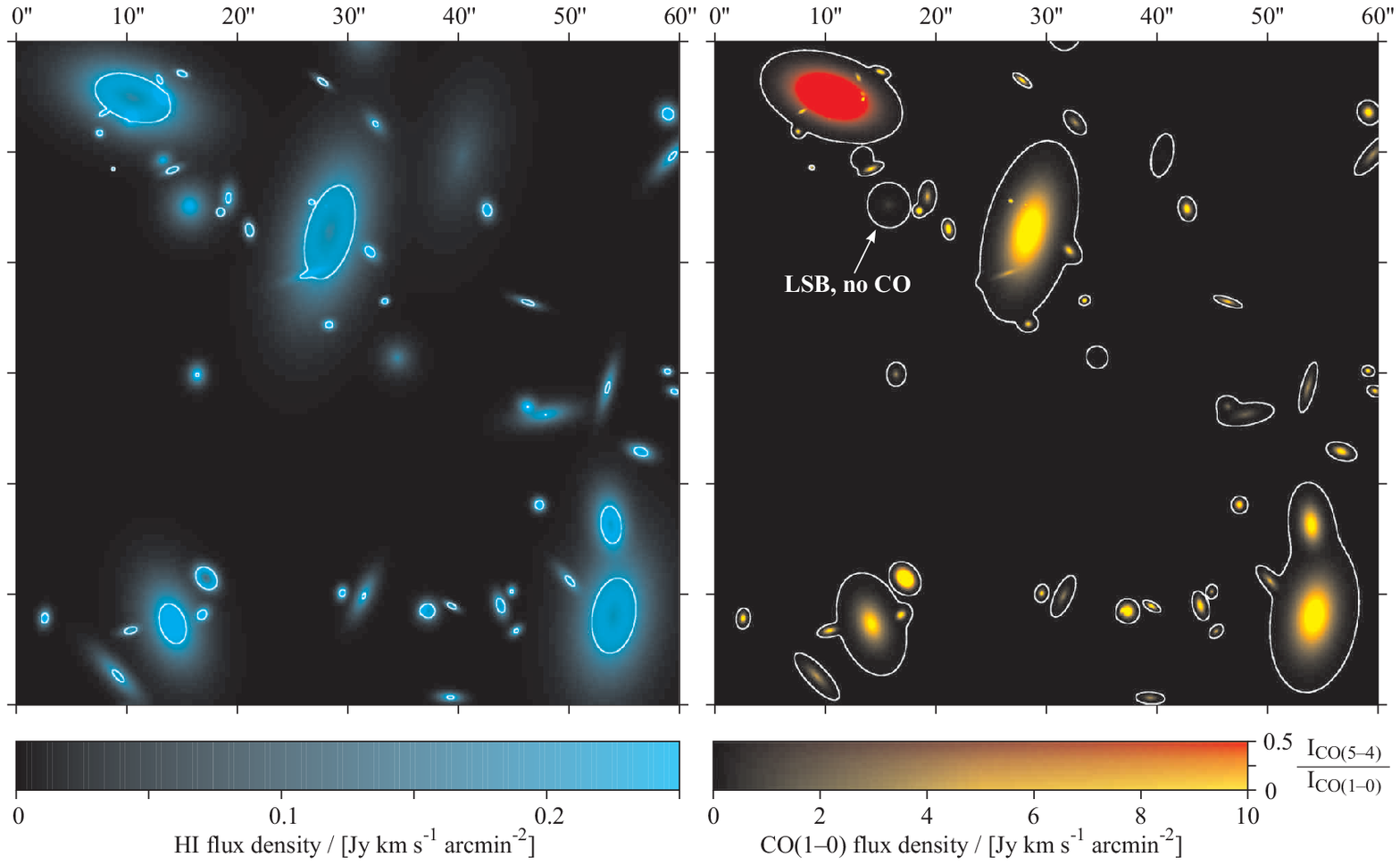}
  \caption{Illustration of the galaxies in the redshift range $z=1.0-1.1$ in a small field of $1\rm\,arcmin^2$. The full field of view of the observing cone is 60,000-times larger than this example. The gradual coloring represents integrated line fluxes per unit solid angle for \ha~(left) and CO(1--0) (right). The different color tones for CO represent the brightness temperature intensity ratio $I_{\rm CO(5-4)}/I_{\rm CO(1-0)}$. The white contours around \ha-sources represent iso-density curves of CO at the 50-percentile level of the full CO density scale and vise versa.}\label{fig_z1}
\end{figure*}

\subsection{Angular sizes}\label{subsection_angular_sizes}

We shall finalize our simulation of line-emitting galaxies by ascribing an angular distribution (flux per unit solid angle) in the sky to each emission line. To this end we assume that, for a fixed galaxy, the angular distributions of the \ha-flux and the different CO-fluxes are proportional to the angular distributions of \ha~and \hm, respectively. We emphasize, however, that the conversion factor between \hm-masses and CO-fluxes depend on the $J$-level of the CO-transition and on the galaxy, such as outlined in Section \ref{subsubsection_cofluxes}.

Given this assumption, we can infer the line flux densities per unit solid angle from the surface densities $\Sigmaha(r)$ and $\Sigmahm(r)$ given in Eqs.~(\ref{eqsigmaha},\ref{eqsigmahm}). We only need to normalize these densities to the respective line fluxes and to replace the scale radius $\rd$ by the apparent scale radius $\rd/\da$, where $\da=(1+z)^{-1}\,\dc$ is the angular diameter distance.

The shapes of the line-emitting regions need to be corrected for the inclinations $i$. If $\qnha$ and $\qnhm$ respectively denote the intrinsic axis ratios of the atomic and molecular gas in galaxies, then the apparent axis ratios are given by \citep[Eq.~(1) in][]{Kannappan2002}
\bea
  \qha^2 & = & \cos^2 i+\qnha^2\,\sin^2 i, \\
  \qhm^2 & = & \cos^2 i+\qnhm^2\,\sin^2 i.
\eea
These relations satisfy $\qha=\qhm=1$, if $i=0{\rm\,deg}$ (face-on), and
$\qha=\qnha$ and $\qhm=\qnhm$, if $i=90{\rm\,deg}$ (edge-on).

We assume that all galaxies have the same values for respectively $\qnha$ and $\qnhm$. In local spiral galaxies, we typically find $\qnha=0.1$, as can be seen from high-resolution maps of the edge-on spiral galaxies NGC 891 and NGC 4565 \citep{Rupen1991}. To our knowledge, no reliable estimate of $\qnhm$ for disk galaxies is available. However, simultaneous CO(1--0) and optical observations revealed that the density of stars in nearby galaxies correlates with the density of molecular gas, when averaged over sufficiently large ($>\rm kcp$) scales \citep{Richmond1986,Leroy2008}, probably as a natural consequence of the formation of stars from molecular gas. Therefore, we shall assume that the intrinsic aspect ratio of molecular gas $\qnhm$ is identical to that of stellar disks. The latter is $\sim0.1$, as can be seen in the sample of 34 nearby edge-on spiral galaxies studied by \cite{Kregel2002}. We therefore adopt $\qnhm=0.1$. We stress that $\qnha=\qnhm$ does not contradict the fact that characteristic scale radii and scale heights of \ha-distributions are generally larger than those of \hm-distributions \citep{Leroy2008}.

The assumption of constant values for $\qnha$ and $\qnhm$ for all galaxies at all redshifts may not be verified at high redshifts due to time-scale arguments. In fact, a significant fraction of the galaxies at $z>5$ may have an age comparable to their dynamical time-scale. Their cold gas distribution might therefore be thicker and less circularly symmetric than the flat gas disks seen today. However, no reliable estimates of $\qnha$ and $\qnhm$ beyond the local Universe are available today.

\newpage

\section{Results}\label{section_results}

The simulated mock observing cone can be accessed on-line as described in Appendix \ref{appendix_access}. This section starts with a graphical illustration of the simulated mock observing cone. As a second step, we analyze the predicted number of galaxies detected in an idealized line-survey with a constant peak flux density limit. Specific predictions for particular surveys with radio and (sub)millimeter telescopes, such as the SKA, the LMT, and ALMA, shall follow in forthcoming studies.

\subsection{Graphical overview}\label{subsection_graphical_overview}

By successively applying the simulation steps described in Section \ref{section_methods}, we have constructed
an observing cone of line emitting galaxies. Fig.~\ref{fig_cone} shows a longitudinal slice of this cone with a thickness of $10\rm~Mpc$. This slice corresponds to a diagonal cut, as illustrated in Fig.~\ref{fig_cone} and has an opening angle of $5.8{\rm\,deg}$. Each pixel inside this slice corresponds to a galaxy. The structure of the cosmic web appears clearly, as well as the increasing filamentarity of this structure with decreasing redshift. The color scales represents the \hm/\ha-mass ratios of the galaxies. We can clearly recognize the pressure-driven cosmic decline of this ratio \citep[see][]{Obreschkow2009c}.

The mock observing cone translates into a virtual sky field when projected onto a sphere using Eqs.~(\ref{eqra},\ref{eqdec}). Fig.~\ref{fig_z1} displays the \ha- and CO-flux densities of galaxies between $z=1$ and $z=1.1$ in a small extract of this virtual sky. The surface densities of the galaxies have been modeled using Eqs.~(\ref{eqsigmaha},\ref{eqsigmahm}) in the manner described in Section \ref{subsection_angular_sizes}. The more massive galaxies in the field reveal ring-like \ha-distributions with CO-rich central regions. By contrast, some of the smaller galaxies with low surface brightness (LSB), have most of their \ha~in the center, with nearly no detectable CO. In general, CO (and hence \hm) is more compact than \ha. All these simulated features compare well
to observed radial distributions (averaged on suitably large scales) of \ha~and CO in nearby galaxies \citep{Leroy2008} as demonstrated in \cite{Obreschkow2009b} and \cite{Obreschkow2009d}. Nevertheless, we emphasize that the idealized nature (e.g.~axial symmetry) of the simulated galaxies misses out a list of structural features found in many real galaxies, such as spiral arms, warps, and lopsidedness.

The coloring of the CO-surface densities in Fig.~\ref{fig_z1} (right) represents the CO(5--4)/CO(1--0) line ratio in terms of brightness temperatures. For normal galaxies, without a particular source of heating, this ratio is much smaller than unity (yellow coloring), however for some galaxies with strong heating by an ongoing SB or AGN, the higher order lines can get excited (red coloring). These mechanisms and our model to assess them are discussed in \cite{Obreschkow2009e}.

\begin{figure*}[t!]
  \includegraphics[width=\textwidth]{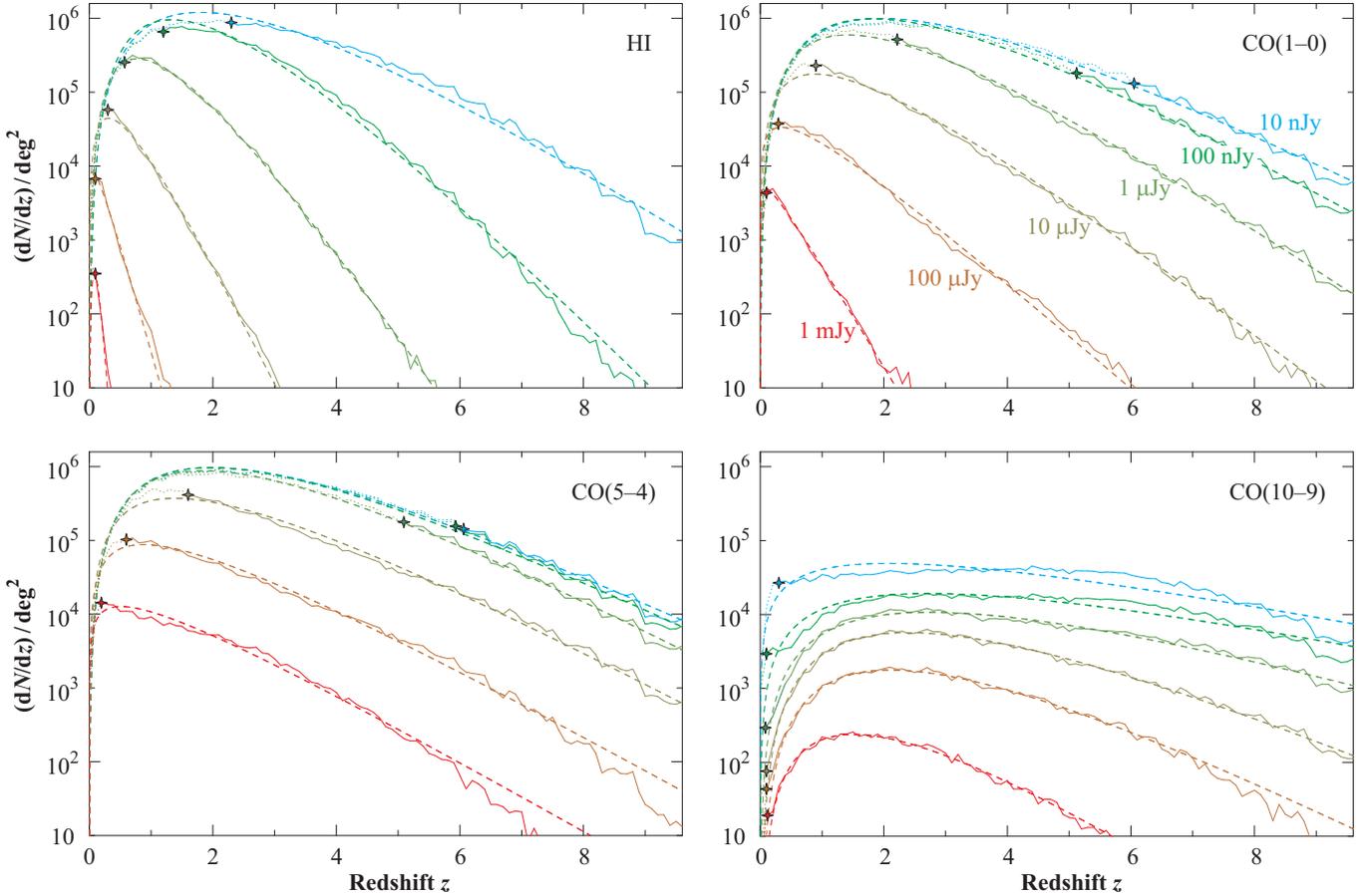}
  \caption{$\dndz$-plots for the emission lines of \ha, CO(1--0), CO(5--4), and CO(10--9) at different peak flux density limits. Solid lines represent the simulated data in the regime where the simulated galaxy sample is complete. Dotted lines represent the cases where the flux density limit is low enough to detect sources in the incomplete parts of the galaxy mass function (see Section \ref{subsection_dndz}). These dotted lines should be considered as lower limits. To increase the readability, the transitions from the complete regime (solid lines) to the incomplete regime (dashed lines) have been marked with stars. All simulated functions (solid and dotted lines) use a redshift bin size of $\Delta z=0.1$ and a sky field of $4\times4\rm~deg^2$, corresponding to a maximal redshift $\zmax=10$. Dashed lines represent analytic fits of Eq.~(\ref{eqdndzfit}). The respective fitting parameters are listed in table \ref{tab_parameters}. Colors correspond to the peak flux density limits shown in the panel for CO(1--0).}\label{fig_dndz}
\end{figure*}

Fig.~\ref{fig_bigmap} in Appendix \ref{appendix_illstration} shows a 3-times larger sky field than Fig.~\ref{fig_z1} at the three redshifts $z\approx1$, $z\approx3$, and $z\approx6$. The progression from $z\approx1$ to $z\approx6$ in Fig.~\ref{fig_bigmap}
reveals two notable features. Firstly, galaxy sizes decrease with redshift. In fact, the angular diameter distances at $z\approx1$ and $z\approx3$ are virtually identical according to the cosmology adopted in this paper (Section \ref{subsection_galaxysim}). Therefore, the galaxy sizes of these two virtual sky maps can be compared directly. The angular diameter distance at $z\approx6$
is about $25\%$ smaller, hence the same physical scales appear slightly oversized. The size evolution of the galaxies reflects the cosmic evolution of the volume/mass-ratio of the dark matter haloes \citep{Gunn1972}. We discussed the impact of this evolution on the surface densities of \ha~and \hm~in \cite{Obreschkow2009d}.

Secondly, the CO(5--4)/CO(1--0) line ratios of CO-detectable galaxies increases with redshift. In fact, at $z\approx6$ no galaxy with a line ratio significantly below unity (i.e.~with yellow coloring) can be seen. This feature relies partially on the compactness of the galaxies, which, according to our model for CO-lines \citep{Obreschkow2009e}, allows an efficient heating by star formation. An additional reason for the absence of low CO(5--4)/CO(1--0) line ratios at $z\gtrsim6$ is that molecular gas in galaxies with no significant star formation and no AGN will be hardly detectable in CO due to its near thermal equilibrium with the CMB.

\subsection{$\dndz$ for a peak flux density limited survey}\label{subsection_dndz}

In the simulated observing cone, we can readily count the number of galaxies per redshift interval with line fluxes above a certain threshold. This $\dndz$-analysis is a key step towards a prediction of the number of sources detectable with any particular telescope and survey strategy. In this section, we focus on the number of sources detected in a peak flux density limited survey, and we restrict the presented results to the \ha, CO(1--0), CO(5--4), and CO(10--9) emission lines. Results for other CO emission lines and/or for integrated flux limited surveys are presented in Appendix \ref{appendix_parameters}.

Fig.~\ref{fig_dndz} shows the $\dndz$-functions for six different peak flux density limits, logarithmically spaced between $1\rm~mJy$ and $10\rm~nJy$. Peak flux densities for each source and emission line are calculated in the way described in Section \ref{subsection_line_profiles}. This method accounts for the different gas distributions, rotation curves, and inclinations of the galaxies. Every source with a peak flux density above the peak flux density limit is considered as detected, while all other sources are considered as non-detected. Different aspects of Fig.~\ref{fig_dndz} will be discussed in detail over the following paragraphs.

\subsubsection{Cosmic variance}

The simulated $\dndz$-functions shown in Fig.~\ref{fig_dndz} (solid and dotted lines) correspond to a bin size of $\Delta z=0.1$ and a sky field of $4\times4\rm~deg^2$, extracted from \emph{one particular} realization of the mock observing cone, that is one random choice of symmetry operations for the replicated simulation boxes (see Section \ref{subsection_mockcone}). The wiggles visible in the simulated $\dndz$-functions are physical. Similar wiggles can be expected for a real sky survey of a sky field of $4\times4\rm~deg^2$ with a redshift bin size of $\Delta z=0.1$. The fact that the amplitude of those wiggles does not decrease as $1/\sqrt{\dndz}$ clearly uncovers the presence of the cosmic large-scale structure, also visible in Fig.~\ref{fig_cone}.

\begin{figure}
  \includegraphics[width=\columnwidth]{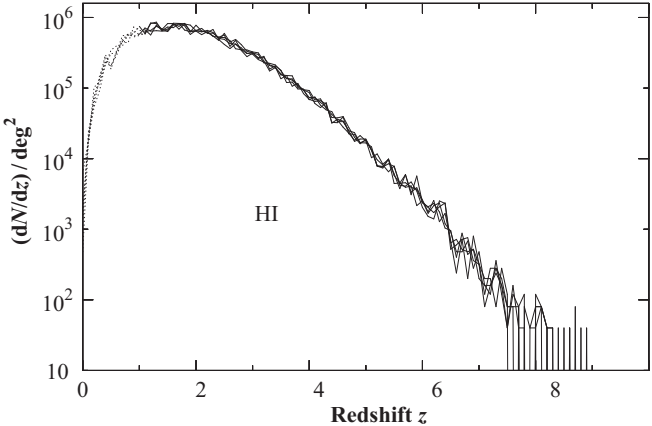}
  \caption{Effects of cosmic variance on a peak flux limited \ha-survey with a flux limit of $1\rm~\mu Jy$. The five lines show the $\dndz$-functions extracted from five distinct random realizations of the mock observing cone (see Section \ref{subsection_mockcone}). Each function uses a bin size of $\Delta z=0.1$ and a small sky field of $1\times1\rm~deg^2$ in order to illustrate the effects of cosmic variance.}\label{fig_dndz_cv}
\end{figure}

To quantify the effects of cosmic variance, we now consider the $\dndz$-functions extracted from five different random realizations of the mock observing cone. Fig.~\ref{fig_dndz_cv} shows the corresponding $\dndz$-functions for a peak flux limited \ha-survey with a peak flux limit of $1\rm~\mu Jy$. Each function uses a bin size of $\Delta z=0.1$ and a small sky field of $1\times1\rm~deg^2$ in order to make the effects of cosmic structure obvious. As a rough estimate the log-scatter between the different $\dndz$-functions is about $0.1\rm~dex$. From this small scatter we conclude that cosmic variance is, in most cases, negligible compared to the uncertainties of the semi-analytic galaxy model.

However, the comoving volume per unit solid angle and unit redshift varies as a function of redshift. Therefore, the scatter due to cosmic variance varies with redshift. It is largest at the lowest redshifts ($z<0.5$), where the comoving surface per unit solid angle is small, and at the highest redshifts ($z>5$), where the radial comoving distance per unit of redshift is small. In these redshift regimes effects of cosmic variance should therefore be estimated, when comparing simulated data to observations.

\subsubsection{Completeness}

The semi-analytic simulation is complete for galaxies with total hydrogen masses (\ha+\hm) above $\sim10^8\,\msun$. Galaxies with smaller hydrogen masses typically sit at the centers of halos with less than 20 particles in the Millennium Simulation, which cannot be reliably identified. Therefore, the number of these low-mass galaxies is heavily underestimated by the simulation. In principle, the sensitivity of a line survey can be high enough that galaxies in this incomplete region of the simulated MF are detected.

In fact, for each emission line and each sensitivity limit, there is a critical redshift $\zc$, below which sources of the incomplete part of the simulated galaxy MF will be detected at a sufficiently high rate that the number of real detections will significantly exceed the number of simulated detections. Therefore, the simulated $\dndz$-functions at $z\leq\zc$ must be considered as lower limits.

We shall here define $\zc$ as the redshift, below which the fraction of sources with hydrogen masses below $10^8\,\msun$ becomes larger than $1\%$. This may seem to be a very low threshold, but we stress that the incomplete part of the simulated hydrogen MF misses a significant fraction of the ``real'' sources and hence the value of $\dndz$ may be underestimated by much more than $1\%$ if $z\leq\zc$. In Fig.~\ref{fig_dndz} the values of $\zc$ have been marked with stars, and the simulated $\dndz$-functions at $z\leq\zc$ have been plotted as dotted lines instead of solid ones.

\subsubsection{Parametrization of the dN/dz-plots}

The simulated $\dndz$-functions can easily be recovered from the on-line database of the sky simulation (see Appendix \ref{appendix_access}). Alternatively, we also approximated the $\dndz$-functions by analytic fits of the form
\be\label{eqdndzfit}
  \frac{{\rm d}N/{\rm d}z}{\rm~deg^2} = 10^{c_1}\cdot z^{c_2}\cdot\exp(-c_3\cdot z),
\ee
where $c_1$, $c_2$, and $c_3$ are free parameters. The best parameters in terms of an rms-minimization are shown in Table \ref{tab_parameters} for various emission lines detected with different limits for the peak flux densities and integrated fluxes. Analytic $\dndz$-functions for intermediate flux limits can be approximately inferred by linearly interpolating the parameters $c_1$, $c_2$, and $c_3$.

\begin{figure*}[t!]
  \includegraphics[width=\textwidth]{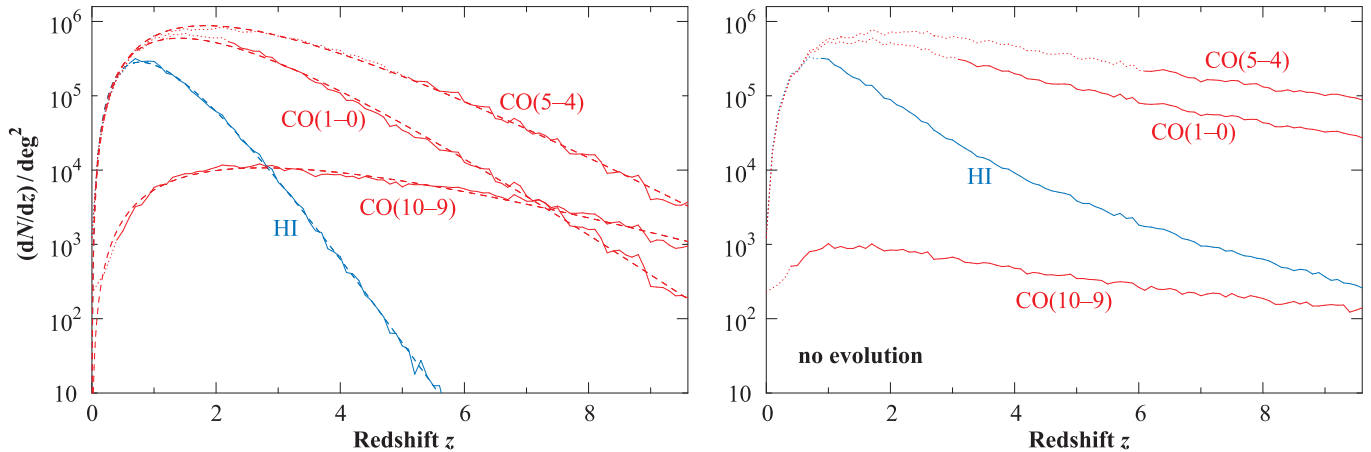}
  \caption{Comparison of the $\dndz$-plots for different emission lines observed with an identical peak flux density limit of $1\rm\,\mu Jy$. The left panel shows our simulation presented in this paper. For comparison, the right panel represents the case of a simulation with no galaxy evolution, as obtained by using only the local galaxy simulation-box for the construction of the mock observing cone. The line types are as explained in Fig.~\ref{fig_dndz}, and colors have been used to distinguish \ha~(blue) from CO (red). All simulated functions (solid and dotted) use a redshift bin size of $\Delta z=0.1$ and a sky field of $4\times4\rm~deg^2$, corresponding to a maximal redshift $\zmax=10$.}\label{fig_dndz2}
\end{figure*}

\subsubsection{Basic conclusions}

An important conclusion from Fig.~\ref{fig_dndz} is that \ha-surveys at high redshift ($z\gtrsim2$) will be difficult compared to CO-surveys. For example, in order to detect the same number of sources at $z\approx4$, an \ha-survey will need to be approximately 10-times more sensitive than a CO(1--0)-survey and over 100-times more sensitive than a CO(5--4)-survey. However, we emphasize that with single dish receivers this effect is partially offset by the fact the at the instantaneous field of view increases as $\lambda^2$, where $\lambda$ is the observed wavelength. This scaling implies that single dish \ha-surveys are likely to yield much more integration time per unit solid angle, which effectively increases the sensitivity. Furthermore, new technologies, such as aperture arrays \citep{Carilli2004b}, are currently being developed to provide an enormous instantaneous field of view and hence a very high effective sensitivity for the detection of \ha.

Fig.~\ref{fig_dndz2} (left) shows a comparison of the simulated $\dndz$-functions for different emission lines observed with an identical peak flux density limit of $1\rm\,\mu Jy$. The flat slope of the $\dndz$-function for CO(10--9) reflects that this line is boosted by SBs, which were more abundant and effective (more compact galaxies) at high $z$ \citep[see][]{Obreschkow2009e}. CO(1--0) reveals the steepest slope of all the CO-lines in the $\dndz$-plot. On one hand, this feature indicates that local galaxies are dominated by low-order excitations of the CO-molecule, consistent with empirical data \citep{Braine1993b}. On the other hand, CO(1--0) becomes nearly invisible in normal (i.e.~no AGN, no SB) galaxies at high redshift ($z>7$) due to a near thermal equilibrium between the molecular gas and the CMB (\citealp{Obreschkow2009e}; see also \citealp{Combes1999,Papadopoulos2000}). The even steeper slope of the $\dndz$-function for \ha~originates from the cosmic decline of the \hm/\ha-ratio in galaxies described in \cite{Obreschkow2009c}.

Fig.~\ref{fig_dndz2} (right) shows the same $\dndz$-functions as Fig.~\ref{fig_dndz2} (left), but for the case of no galaxy evolution. These functions were obtained by constructing a mock observing cone using only the simulation-box at $z=0$. The comparison of Fig.~\ref{fig_dndz2} (left) to Fig.~\ref{fig_dndz2} (right) reveals that \ha~at high redshifts will be much harder to detect than predicted by a no-evolution model. Qualitatively, the same conclusion applies to low-order CO-emission lines, but the effect is less significant. In contrast, our simulation predicts that the high-order CO-emission lines will be easier to detect than suggested by a no-evolution model, since these lines will be strongly boosted by SBs at high redshift.

\section{Discussion}\label{section_discussion}

\subsection{Limitations of the galaxy simulation}\label{subsection_limgen}

Our simulation is inevitably bound to the $\Lambda$CDM cosmology with the cosmological parameters given in Section \ref{section_methods}. The empirical uncertainty of these parameters may be a source of systematic errors in our predictions. To analyze the errors associated with the uncertainty of the Hubble constant, we can study the change of our predictions in the linear expansion\footnote{In the simulation, both masses and lengths scale as $\h^{-1}$ \citep{Springel2005}.} of $\h$. This analysis shows that varying $\h$ between $0.6$ and $0.8$ does not significantly affect the $\dndz$-functions, i.e.~not more than a factor 2. Additionally, \cite{Wang2008} showed that the lower value for the fluctuation amplitude $\sigma_8$ found by \mbox{WMAP-3} compared to the value used in the Millennium simulation is almost entirely compensated by an increase in halo bias. Caution should nevertheless be applied when relying on predictions from a single cosmological model.

An additional limitation of the Millennium Simulation is the mass resolution of $8.6\cdot10^8\,\msun$ per particle. This mass scale sets the completeness limit in our hydrogen simulation to $\mha+\mhm\approx10^8\,\msun$ (Section \ref{subsection_dndz}; \citealp{Obreschkow2009b}). Moreover, galaxies with $\mha+\mhm\lesssim10^9\,\msun$ normally sit a the centers of dark matter halos with poorly resolved merger histories. Therefore, their properties may not have converged in the semi-analytic simulation \citep{Obreschkow2009b,Croton2006}.

A long list of limitations associated with the semi-analytic galaxy simulation and our post-processing to assign extended \ha- and \hm-properties has been considered by \cite{Obreschkow2009b}. The bottom line of this discussion is that, at $z\gtrsim5$, the simulation becomes very uncertain because the geometries and matter content of regular galaxies are virtually unconstrained from an empirical viewpoint. The young age and short merger intervals of these galaxies may, in fact, have caused them to deviate substantially from the simplistic disk-gas model. At $z\lesssim5$, the predictions of our \ha- and \hm-properties are more certain, as they are consistent with available observations. For example, two measurements of CO(2--1)-line emission in regular galaxies at $z\approx1.5$ \citep{Daddi2008} are consistent with the \hm-MF at this redshift \citep{Obreschkow2009c}. Furthermore, the predicted comoving space density of \hm~evolves proportionally to the observed space density of star formation \citep[e.g.][]{Hopkins2007} within a factor 2 out to at least $z=3$. At $z=0$, the simulated \ha-MF and CO(1--0)-luminosity function are consistent with the observations of \cite{Zwaan2005} and \cite{Keres2003}. Additionally, the local sizes and line widths of \ha~and CO match the local observations \citep[][and references therein]{Obreschkow2009b}.

We shall now highlight some specific limitations associated with the emission lines considered in this paper.

\subsection{Limitations specific to the \ha-line}\label{subsection_limhi}

We emphasize that at high redshift, the simulated cosmic \ha-space density $\Omegaha$ falls below the inferences from Lyman-$\alpha$ absorption against distant QSOs by a factor $\sim2$. As mentioned in \cite{Obreschkow2009c}, this could reflect a serious limitations of the semi-analytic models implied by the treatment of all cold hydrogen (\ha+\hm) as a single phase. Consequently, our $\dndz$-predictions for \ha~could be slightly pessimistic. If we believe the empirical estimations of $\Omegaha$, the offset of our \ha-masses by a factor $\sim2$ can be readily accounted for by artificially decreasing the flux limit of the simulated survey by a factor 2. For typical \ha-surveys in the redshift range $z=0.5-10$, this would increase the number of detectable sources by a factor $2-4$.

We have limited our predictions for \ha~to \ha-emission from galaxies. However, in the EoR, the IGM was not completely ionized and therefore acted as an additional source of \ha-emission or -absorption \citep{Iliev2002}. It may therefore be necessary to analyze the implications of intergalactic \ha~on the detectability of galactic \ha~at $z\gtrsim6$ \citep{Becker2001}. On a theoretical level, such an analysis could result from combining the simulation presented in this paper with a simulation of the EoR \citep[e.g.][]{Baek2009,Santos2008}.

\subsection{Limitations specific to the CO-lines}\label{subsection_limco}

The main discussion of these limitations is given in \cite{Obreschkow2009c}, where we introduced our model for the conversion between \hm~and CO. The most serious sources of uncertainty appear to be the heating of molecular gas by SBs and AGNs, the overlap of molecular clouds at high redshift, and the possible presence of nuclear molecular disks in high-redshift galaxies. By contrast, the often discussed effects of the CMB and the cosmic evolution of the metallicity seem relatively well understood today. Overall, the uncertainty in the predicted CO-luminosities increases with redshift and with the $J$-level of the CO-transition.

The highest uncertainties, i.e.~those for the higher order CO-lines at high redshift, can be close to a factor 10. The $\dndz$-functions in this regime are therefore expected to deviate significantly from our predictions. Such deviations will uncover much of the physics of CO-line emission. In fact, in \cite{Obreschkow2009c} we have explained in detail how different deviations of the CO-luminosity functions from our predictions can be translated into physical interpretations.

\subsection{Is the simulation large enough to probe cosmic structure?}\label{subsection_structure}

The largest coherence-scale of our sky simulation is defined by the size of the periodic simulation box of the underlying dark matter simulation \citep[Millennium Simulation,][]{Springel2005}. The side length of this box is $\simbox=500\,\h^{-1}\,\rm Mpc$, which sets the smallest extractable wave number to $k=2\pi/\simbox\approx0.013\,\h$. This value is comparable to the wave number of the first peak in the CDM power spectrum \citep[e.g.][]{Springel2005}. Therefore, the presented simulation allows us to study the power spectrum of \ha- and CO-lines and to extract the baryon acoustic oscillations (BAOs); however the position and the amplitude of the first peak of the BAOs will be very poorly constrained.

By contrast, the SKA will have the potential to improve on present measurements of the baryonic power spectrum by at least an order of magnitude in amplitude, and it will detect power in spatial frequencies far below the first acoustic peak. Such a detection could set a primordial constraint on cosmological parameters, especially on the equation of state of dark energy \citep{Blake2004,Abdalla2008}. Therefore, a simulation of such a detection is regarded as a necessary step in designing the SKA. Yet, this requirement represents a major challenge since no current simulation of cosmic structure is large enough to accurately follow the largest acoustic oscillations, while simultaneously resolving structures small enough to allow the assembly of typical galaxies.

A circumvention of this numerical predicament could result from merging two simulations with different length-scales \citep[see~e.g.][]{Angulo2008}. For example, we could adopt the Horizon-$4\pi$ dark matter simulation \citep{Prunet2008,Teyssier2008}, which has a giant box side length of $\simbox=2\,\h^{-1}\,{\rm Gpc}$, yet 10-times less mass resolution than the Millennium Simulation. Each dark matter halo of the Horizon-$4\pi$ Simulation could then be populated with the resolved dark matter substructure and the galaxies contained in comparable haloes of the Millennium Simulation.

\section{Conclusion}\label{section_conclusion}

With this paper we release a simulation of the \ha-emission line and the first ten $\rm ^{12}C^{16}O$-emission lines of galaxies in a sky field with a comoving diameter of $500\,\h^{-1}\rm\,Mpc$. The actual field-of-view depends on the (user-defined) maximal redshift $\zmax$ according to the relation displayed in Fig.~\ref{fig_fov} (see also Eq.~\ref{eqphiboxes}). This simulation represents the first quantitative attempt to compare the detectability of \ha~and CO at high redshift. Despite the limitations and uncertainties of this simulation (Section \ref{section_discussion}), its underlying galaxy simulation is nonetheless consistent with currently available observations \citep[see][]{Obreschkow2009b,Obreschkow2009c,Obreschkow2009e}.

While this paper focussed on the simulation techniques and directly accessible results, the list of possible applications of the presented simulation is extensive. Some examples are:
\begin{enumerate}
  \item a $\dndz$-analysis for particular surveys with the SKA, the LMT, and ALMA or their pathfinders;
  \item a combined study of \ha-emission from galaxies and \ha-emission from the IGM during the EoR \citep{Santos2008} to quantify confusion issues;
  \item an optimization of the survey time allocated to different ALMA bands based on the CO-line ratios predicted at various redshifts;
  \item first quantitative predictions of the SKA's and ALMA's abilities to probe the galaxy power spectrum;
  \item predictions of the absorption signatures of \ha~and CO against distant QSOs;
  \item a study of line stacking experiments at redshifts where the detection of individual galaxies becomes impossible. \\
\end{enumerate}

Such predictions can assist the design and optimized use of telescopes, such as the LMT, the SKA, ALMA, and their pathfinders. Moreover, in light of forthcoming observations with those future instruments, the predictions made \emph{prior} to these observations are the safest and perhaps the only way to test the predictive power of our current theories. This feature accentuates the necessity of extensive predictions, whether they will be verified or not by the empirical data.

\acknowledgments
This effort/activity is supported by the European Community Framework Programme 6, Square Kilometre Array Design Studies (SKADS), contract no 011938. The Millennium Simulation databases and the web application providing online access to them were constructed as part of the activities of the German Astrophysical Virtual Observatory.


\begin{thebibliography}{61}
\expandafter\ifx\csname natexlab\endcsname\relax\def\natexlab#1{#1}\fi

\bibitem[{{Abdalla} {et~al.}(2009){Abdalla}, {Blake} \&
  {Rawlings}}]{Abdalla2008}
{Abdalla} F., {Blake} C., {Rawlings} S., 2009, \mnras, submitted

\bibitem[{{Abdalla} \& {Rawlings}(2005)}]{Abdalla2005}
{Abdalla} F.~B., {Rawlings} S., 2005, \mnras, 360, 27

\bibitem[{{Angulo} {et~al.}(2008){Angulo}, {Baugh}, {Frenk} \&
  {Lacey}}]{Angulo2008}
{Angulo} R.~E., {Baugh} C.~M., {Frenk} C.~S., {Lacey} C.~G., 2008, \mnras, 383,
  755

\bibitem[{{Baek} {et~al.}(2009){Baek}, {di Matteo}, {Semelin}, {Combes} \&
  {Revaz}}]{Baek2009}
{Baek} S., {di Matteo} P., {Semelin} B., {Combes} F., {Revaz} Y., 2009, \aap,
  495, 389

\bibitem[{{Becker} {et~al.}(2001){Becker}, {Fan}, {White}, {Strauss},
  {Narayanan}, {Lupton}, {Gunn}, {Annis}, {Bahcall}, {Brinkmann}, {Connolly},
  {Csabai}, {Czarapata}, {Doi}, {Heckman}, {Hennessy}, {Ivezi{\'c}}, {Knapp},
  {Lamb}, {McKay}, {Munn}, {Nash}, {Nichol}, {Pier}, {Richards}, {Schneider},
  {Stoughton}, {Szalay}, {Thakar} \& {York}}]{Becker2001}
{Becker} R.~H., {et~al.}, 2001, \aj, 122, 2850

\bibitem[{{Binney} \& {Merrifield}(1998)}]{Binney1998}
{Binney} J., {Merrifield} M., 1998, {Galactic astronomy}. Princeton University
  Press

\bibitem[{{Blain} {et~al.}(2000){Blain}, {Frayer}, {Bock} \&
  {Scoville}}]{Blain2000}
{Blain} A.~W., {Frayer} D.~T., {Bock} J.~J., {Scoville} N.~Z., 2000, \mnras,
  313, 559

\bibitem[{{Blaizot} {et~al.}(2005){Blaizot}, {Wadadekar}, {Guiderdoni},
  {Colombi}, {Bertin}, {Bouchet}, {Devriendt} \& {Hatton}}]{Blaizot2005}
{Blaizot} J., {Wadadekar} Y., {Guiderdoni} B., {Colombi} S.~T., {Bertin} E.,
  {Bouchet} F.~R., {Devriendt} J.~E.~G., {Hatton} S., 2005, \mnras, 360, 159

\bibitem[{{Blake} {et~al.}(2004){Blake}, {Abdalla}, {Bridle} \&
  {Rawlings}}]{Blake2004}
{Blake} C.~A., {Abdalla} F.~B., {Bridle} S.~L., {Rawlings} S., 2004, New
  Astronomy Review, 48, 1063

\bibitem[{{Blitz} \& {Rosolowsky}(2006)}]{Blitz2006}
{Blitz} L., {Rosolowsky} E., 2006, \apj, 650, 933

\bibitem[{{Boomsma} {et~al.}(2002){Boomsma}, {van der Hulst}, {Oosterloo} \&
  {Sancisi}}]{Boomsma2002}
{Boomsma} R., {van der Hulst} J.~M., {Oosterloo} T.~A., {Sancisi} R., 2002, in
  Bulletin of the American Astronomical Society, Vol.~34, p. 708

\bibitem[{{Braine} {et~al.}(1993){Braine}, {Combes}, {Casoli}, {Dupraz},
  {Gerin}, {Klein}, {Wielebinski} \& {Brouillet}}]{Braine1993b}
{Braine} J., {Combes} F., {Casoli} F., {Dupraz} C., {Gerin} M., {Klein} U.,
  {Wielebinski} R., {Brouillet} N., 1993, \aaps, 97, 887

\bibitem[{{Carilli} \& {Blain}(2002)}]{Carilli2002b}
{Carilli} C.~L., {Blain} A.~W., 2002, \apj, 569, 605

\bibitem[{{Carilli} \& {Rawlings}(2004)}]{Carilli2004b}
{Carilli} C.~L., {Rawlings} S., 2004, New Astronomy Review, 48, 979

\bibitem[{{Catinella} {et~al.}(2008){Catinella}, {Haynes}, {Giovanelli},
  {Gardner} \& {Connolly}}]{Catinella2008}
{Catinella} B., {Haynes} M.~P., {Giovanelli} R., {Gardner} J.~P., {Connolly}
  A.~J., 2008, \apjl, 685, L13

\bibitem[{{Cole} {et~al.}(2001){Cole}, {Norberg}, {Baugh}, {Frenk},
  {Bland-Hawthorn}, {Bridges}, {Cannon}, {Colless}, {Collins}, {Couch},
  {Cross}, {Dalton}, {De Propris}, {Driver}, {Efstathiou}, {Ellis},
  {Glazebrook}, {Jackson}, {Lahav}, {Lewis}, {Lumsden}, {Maddox}, {Madgwick},
  {Peacock}, {Peterson}, {Sutherland} \& {Taylor}}]{Cole2001}
{Cole} S., {et~al.}, 2001, \mnras, 326, 255

\bibitem[{{Combes} {et~al.}(1999){Combes}, {Maoli} \& {Omont}}]{Combes1999}
{Combes} F., {Maoli} R., {Omont} A., 1999, \aap, 345, 369

\bibitem[{{Croton} {et~al.}(2006){Croton}, {Springel}, {White}, {De Lucia},
  {Frenk}, {Gao}, {Jenkins}, {Kauffmann}, {Navarro} \& {Yoshida}}]{Croton2006}
{Croton} D.~J., {et~al.}, 2006, \mnras, 365, 11

\bibitem[{{Daddi} {et~al.}(2008){Daddi}, {Dannerbauer}, {Elbaz}, {Dickinson},
  {Morrison}, {Stern} \& {Ravindranath}}]{Daddi2008}
{Daddi} E., {Dannerbauer} H., {Elbaz} D., {Dickinson} M., {Morrison} G.,
  {Stern} D., {Ravindranath} S., 2008, \apjl, 673, L21

\bibitem[{{De Lucia} \& {Blaizot}(2007)}]{DeLucia2007}
{De Lucia} G., {Blaizot} J., 2007, \mnras, 375, 2

\bibitem[{{Ferri{\`e}re}(2001)}]{Ferriere2001}
{Ferri{\`e}re} K.~M., 2001, Reviews of Modern Physics, 73, 1031

\bibitem[{{F{\"o}rster Schreiber} {et~al.}(2006){F{\"o}rster Schreiber},
  {Genzel}, {Lehnert}, {Bouch{\'e}}, {Verma}, {Erb}, {Shapley}, {Steidel},
  {Davies}, {Lutz}, {Nesvadba}, {Tacconi}, {Eisenhauer}, {Abuter}, {Gilbert},
  {Gillessen} \& {Sternberg}}]{ForsterSchreiber2006}
{F{\"o}rster Schreiber} N.~M., {et~al.}, 2006, \apj, 645, 1062

\bibitem[{{Genzel} {et~al.}(2008){Genzel}, {Burkert}, {Bouch{\'e}}, {Cresci},
  {F{\"o}rster Schreiber}, {Shapley}, {Shapiro}, {Tacconi}, {Buschkamp},
  {Cimatti}, {Daddi}, {Davies}, {Eisenhauer}, {Erb}, {Genel}, {Gerhard},
  {Hicks}, {Lutz}, {Naab}, {Ott}, {Rabien}, {Renzini}, {Steidel}, {Sternberg}
  \& {Lilly}}]{Genzel2008}
{Genzel} R., {et~al.}, 2008, \apj, 687, 59

\bibitem[{{Giovanelli} {et~al.}(1997){Giovanelli}, {Haynes}, {da Costa},
  {Freudling}, {Salzer} \& {Wegner}}]{Giovanelli1997}
{Giovanelli} R., {Haynes} M.~P., {da Costa} L.~N., {Freudling} W., {Salzer}
  J.~J., {Wegner} G., 1997, \apjl, 477, L1

\bibitem[{{Greve} {et~al.}(2009){Greve}, {Papadopoulos}, {Gao} \&
  {Radford}}]{Greve2009}
{Greve} T.~R., {Papadopoulos} P.~P., {Gao} Y., {Radford} S.~J.~E., 2009, \apj,
  692, 1432

\bibitem[{{Greve} \& {Sommer-Larsen}(2008)}]{Greve2008}
{Greve} T.~R., {Sommer-Larsen} J., 2008, \aap, 480, 335

\bibitem[{{Gunn} \& {Gott}(1972)}]{Gunn1972}
{Gunn} J.~E., {Gott} J.~R.~I., 1972, \apj, 176, 1

\bibitem[{{H{\"a}ring} \& {Rix}(2004)}]{Haering2004}
{H{\"a}ring} N., {Rix} H.-W., 2004, \apjl, 604, L89

\bibitem[{{Hopkins}(2007)}]{Hopkins2007}
{Hopkins} A.~M., 2007, in Astronomical Society of the Pacific Conference
  Series, Vol. 380, Deepest Astronomical Surveys, {Afonso} J., {Ferguson}
  H.~C., {Mobasher} B., {Norris} R., eds., p. 423

\bibitem[{{Huang} {et~al.}(2003){Huang}, {Glazebrook}, {Cowie} \&
  {Tinney}}]{Huang2003}
{Huang} J.-S., {Glazebrook} K., {Cowie} L.~L., {Tinney} C., 2003, \apj, 584,
  203

\bibitem[{{Iliev} {et~al.}(2002){Iliev}, {Shapiro}, {Ferrara} \&
  {Martel}}]{Iliev2002}
{Iliev} I.~T., {Shapiro} P.~R., {Ferrara} A., {Martel} H., 2002, \apjl, 572,
  L123

\bibitem[{{Kannappan} {et~al.}(2002){Kannappan}, {Fabricant} \&
  {Franx}}]{Kannappan2002}
{Kannappan} S.~J., {Fabricant} D.~G., {Franx} M., 2002, \aj, 123, 2358

\bibitem[{{Keres} {et~al.}(2003){Keres}, {Yun} \& {Young}}]{Keres2003}
{Keres} D., {Yun} M.~S., {Young} J.~S., 2003, \apj, 582, 659

\bibitem[{{Kregel} {et~al.}(2002){Kregel}, {van der Kruit} \& {de
  Grijs}}]{Kregel2002}
{Kregel} M., {van der Kruit} P.~C., {de Grijs} R., 2002, \mnras, 334, 646

\bibitem[{{Lah} {et~al.}(2007){Lah}, {Chengalur}, {Briggs}, {Colless}, {de
  Propris}, {Pracy}, {de Blok}, {Fujita}, {Ajiki}, {Shioya}, {Nagao},
  {Murayama}, {Taniguchi}, {Yagi} \& {Okamura}}]{Lah2007}
{Lah} P., {et~al.}, 2007, \mnras, 376, 1357

\bibitem[{{Leroy} {et~al.}(2008){Leroy}, {Walter}, {Brinks}, {Bigiel}, {de
  Blok}, {Madore} \& {Thornley}}]{Leroy2008}
{Leroy} A.~K., {Walter} F., {Brinks} E., {Bigiel} F., {de Blok} W.~J.~G.,
  {Madore} B., {Thornley} M.~D., 2008, \aj, 136, 2782

\bibitem[{{Meyer} {et~al.}(2004){Meyer}, {Zwaan}, {Webster}, {Staveley-Smith},
  {Ryan-Weber}, {Drinkwater}, {Barnes}, {Howlett}, {Kilborn}, {Stevens},
  {Waugh}, {Pierce}, {Bhathal}, {de Blok}, {Disney}, {Ekers}, {Freeman},
  {Garcia}, {Gibson}, {Harnett}, {Henning}, {Jerjen}, {Kesteven}, {Knezek},
  {Koribalski}, {Mader}, {Marquarding}, {Minchin}, {O'Brien}, {Oosterloo},
  {Price}, {Putman}, {Ryder}, {Sadler}, {Stewart}, {Stootman} \&
  {Wright}}]{Meyer2004}
{Meyer} M.~J., {et~al.}, 2004, \mnras, 350, 1195

\bibitem[{{Norberg} {et~al.}(2002){Norberg}, {Cole}, {Baugh}, {Frenk},
  {Baldry}, {Bland-Hawthorn}, {Bridges}, {Cannon}, {Colless}, {Collins},
  {Couch}, {Cross}, {Dalton}, {De Propris}, {Driver}, {Efstathiou}, {Ellis},
  {Glazebrook}, {Jackson}, {Lahav}, {Lewis}, {Lumsden}, {Maddox}, {Madgwick},
  {Peacock}, {Peterson}, {Sutherland} \& {Taylor}}]{Norberg2002}
{Norberg} P., {et~al.}, 2002, \mnras, 336, 907

\bibitem[{{Noterdaeme} {et~al.}(2008){Noterdaeme}, {Ledoux}, {Petitjean} \&
  {Srianand}}]{Noterdaeme2008}
{Noterdaeme} P., {Ledoux} C., {Petitjean} P., {Srianand} R., 2008, \aap, 481,
  327

\bibitem[{{Obreschkow} {et~al.}(2009{\natexlab{a}}){Obreschkow}, {Croton},
  {DeLucia}, {Khochfar} \& {Rawlings}}]{Obreschkow2009b}
{Obreschkow} D., {Croton} D., {DeLucia} G., {Khochfar} S., {Rawlings} S.,
  2009{\natexlab{a}}, \apj, 698, 1467

\bibitem[{{Obreschkow} {et~al.}(2009{\natexlab{b}}){Obreschkow}, {Heywood},
  {Kl{\"o}ckner} \& {Rawlings}}]{Obreschkow2009e}
{Obreschkow} D., {Heywood} I., {Kl{\"o}ckner} H.-R., {Rawlings} S.,
  2009{\natexlab{b}}, \apj, 702, 1321

\bibitem[{{Obreschkow} \& {Rawlings}(2009{\natexlab{a}})}]{Obreschkow2009d}
{Obreschkow} D., {Rawlings} S., 2009{\natexlab{a}}, \mnras, 400, 665

\bibitem[{{Obreschkow} \& {Rawlings}(2009{\natexlab{b}})}]{Obreschkow2009c}
---, 2009{\natexlab{b}}, \apjl, 696, L129

\bibitem[{{Papadopoulos} {et~al.}(2000){Papadopoulos}, {R{\"o}ttgering}, {van
  der Werf}, {Guilloteau}, {Omont}, {van Breugel} \&
  {Tilanus}}]{Papadopoulos2000}
{Papadopoulos} P.~P., {R{\"o}ttgering} H.~J.~A., {van der Werf} P.~P.,
  {Guilloteau} S., {Omont} A., {van Breugel} W.~J.~M., {Tilanus} R.~P.~J.,
  2000, \apj, 528, 626

\bibitem[{{Prochaska} {et~al.}(2005){Prochaska}, {Herbert-Fort} \&
  {Wolfe}}]{Prochaska2005}
{Prochaska} J.~X., {Herbert-Fort} S., {Wolfe} A.~M., 2005, \apj, 635, 123

\bibitem[{{Prunet} {et~al.}(2008){Prunet}, {Pichon}, {Aubert}, {Pogosyan},
  {Teyssier} \& {Gottloeber}}]{Prunet2008}
{Prunet} S., {Pichon} C., {Aubert} D., {Pogosyan} D., {Teyssier} R.,
  {Gottloeber} S., 2008, \apjs, 178, 179

\bibitem[{{Richmond} \& {Knapp}(1986)}]{Richmond1986}
{Richmond} M.~W., {Knapp} G.~R., 1986, \aj, 91, 517

\bibitem[{{Rupen}(1991)}]{Rupen1991}
{Rupen} M.~P., 1991, \aj, 102, 48

\bibitem[{{Santos} {et~al.}(2008){Santos}, {Amblard}, {Pritchard}, {Trac},
  {Cen} \& {Cooray}}]{Santos2008}
{Santos} M.~G., {Amblard} A., {Pritchard} J., {Trac} H., {Cen} R., {Cooray} A.,
  2008, \apj, 689, 1

\bibitem[{{Springel} {et~al.}(2005){Springel}, {White}, {Jenkins}, {Frenk},
  {Yoshida}, {Gao}, {Navarro}, {Thacker}, {Croton}, {Helly}, {Peacock}, {Cole},
  {Thomas}, {Couchman}, {Evrard}, {Colberg} \& {Pearce}}]{Springel2005}
{Springel} V., {et~al.}, 2005, \nat, 435, 629

\bibitem[{{Tacconi} {et~al.}(2006){Tacconi}, {Neri}, {Chapman}, {Genzel},
  {Smail}, {Ivison}, {Bertoldi}, {Blain}, {Cox}, {Greve} \&
  {Omont}}]{Tacconi2006}
{Tacconi} L.~J., {et~al.}, 2006, \apj, 640, 228

\bibitem[{{Teyssier} {et~al.}(2008){Teyssier}, {Pires}, {Prunet}, {Aubert},
  {Pichon}, {Amara}, {Benabed}, {Colombi}, {Refregier} \&
  {Starck}}]{Teyssier2008}
{Teyssier} R., {et~al.}, 2008, ArXiv e-prints

\bibitem[{{Tremonti} {et~al.}(2004){Tremonti}, {Heckman}, {Kauffmann},
  {Brinchmann}, {Charlot}, {White}, {Seibert}, {Peng}, {Schlegel}, {Uomoto},
  {Fukugita} \& {Brinkmann}}]{Tremonti2004}
{Tremonti} C.~A., {et~al.}, 2004, \apj, 613, 898

\bibitem[{{Verheijen} {et~al.}(2007){Verheijen}, {van Gorkom}, {Szomoru},
  {Dwarakanath}, {Poggianti} \& {Schiminovich}}]{Verheijen2007}
{Verheijen} M., {van Gorkom} J.~H., {Szomoru} A., {Dwarakanath} K.~S.,
  {Poggianti} B.~M., {Schiminovich} D., 2007, \apjl, 668, L9

\bibitem[{{Wall}(2006)}]{Wall2006}
{Wall} W.~F., 2006, Rev. Mex. Astron. Astrofis., 42, 117

\bibitem[{{Walter} {et~al.}(2004){Walter}, {Carilli}, {Bertoldi}, {Menten},
  {Cox}, {Lo}, {Fan} \& {Strauss}}]{Walter2004}
{Walter} F., {Carilli} C., {Bertoldi} F., {Menten} K., {Cox} P., {Lo} K.~Y.,
  {Fan} X., {Strauss} M.~A., 2004, \apjl, 615, L17

\bibitem[{{Wang} {et~al.}(2008){Wang}, {De Lucia}, {Kitzbichler} \&
  {White}}]{Wang2008}
{Wang} J., {De Lucia} G., {Kitzbichler} M.~G., {White} S.~D.~M., 2008, \mnras,
  384, 1301

\bibitem[{{Weiss} {et~al.}(2007){Weiss}, {Downes}, {Neri}, {Walter}, {Henkel},
  {Wilner}, {Wagg} \& {Wiklind}}]{Weiss2007}
{Weiss} A., {Downes} D., {Neri} R., {Walter} F., {Henkel} C., {Wilner} D.~J.,
  {Wagg} J., {Wiklind} T., 2007, \aap, 467, 955

\bibitem[{{Wilman} {et~al.}(2008){Wilman}, {Miller}, {Jarvis}, {Mauch},
  {Levrier}, {Abdalla}, {Rawlings}, {Kl{\"o}ckner}, {Obreschkow}, {Olteanu} \&
  {Young}}]{Wilman2008}
{Wilman} R.~J., {et~al.}, 2008, \mnras, 388, 1335

\bibitem[{{Young}(2002)}]{Young2002}
{Young} L.~M., 2002, \aj, 124, 788

\bibitem[{{Zwaan} {et~al.}(2005){Zwaan}, {Meyer}, {Staveley-Smith} \&
  {Webster}}]{Zwaan2005}
{Zwaan} M.~A., {Meyer} M.~J., {Staveley-Smith} L., {Webster} R.~L., 2005,
  \mnras, 359, L30

\end{thebibliography}

\newpage

\appendix

\section{A. How to recover line profiles from catalog properties}\label{appendix_recovery}

\begin{figure}[h]
\centering
  \includegraphics[width=8.5cm]{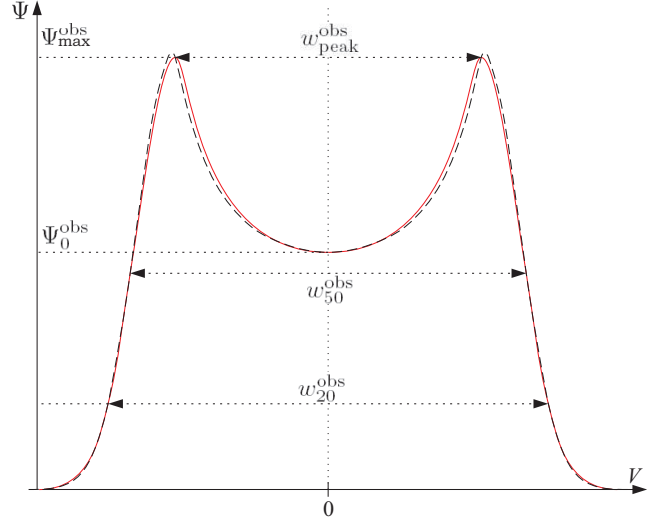}
  \caption{Comparison of a simulated normalized emission line $\Psi(V)$ (solid line) with the emission line $\Psi_{\rm approx}(V)$ (dashed line), recovered from the five parameters $\Psi^{\rm obs}_0$, $\Psi^{\rm obs}_{\rm max}$, $\wobs{peak}$, $\wobs{50}$, and $\wobs{20}$.}\label{fig_recover_profile}
\end{figure}

In Section \ref{subsection_line_profiles}, the normalized profiles of the \ha- and CO-emission lines have been parameterized using the five parameters $\Psi^{\rm obs}_0$, $\Psi^{\rm obs}_{\rm max}$, $\wobs{peak}$, $\wobs{50}$, and $\wobs{20}$ (see Fig.~\ref{fig_recover_profile}). From these parameters, the original normalized velocity profiles $\Psi(V)$ can be approximately recovered using the analytic function
\be\label{eqpsiapprox}
  \Psi_{\rm approx}(V) = \left\{
    \begin{array}{ll}
       \Psi^{\rm obs}_{\rm max}\,\exp\!\left[k_1\,(\abs{V}-k_3)^{k_2}\right]\quad & {\rm if~}\abs{V}\geq\wobs{peak}/2, \\
       k_5\,(k_4-V^2)^{-0.5} & {\rm if~}\abs{V}<\wobs{peak}/2{\rm~and~}\Psi^{\rm obs}_{\rm max}>\Psi^{\rm obs}_0, \\
       \Psi^{\rm obs}_0 & {\rm if~}\abs{V}<\wobs{peak}/2{\rm~and~}\Psi^{\rm obs}_{\rm max}=\Psi^{\rm obs}_0,
    \end{array}
  \right.
\ee
where $k_i$, $i\in\{1,...,5\}$, are free parameters. Eq.~(\ref{eqpsiapprox}) combines the the functional form of Eq.~(42) in \cite{Obreschkow2009b} for the center of the emission line with exponential tails. The five parameters $k_i$ can be inferred from the parameters $\Psi^{\rm obs}_0$, $\Psi^{\rm obs}_{\rm max}$, $\wobs{peak}$, $\wobs{50}$, and $\wobs{20}$. The analytic solution is
\bea
  k_1 & = & -0.693\cdot2.322^{^{^{\left[\frac{\ln(\wobs{50}-\wobs{peak})-\ln{2}}{\ln(\wobs{50}-\wobs{peak})-\ln(\wobs{20}-\wobs{peak})}\right]}}}, \\
  k_2 & = & \frac{0.842}{\ln(\wobs{20}-\wobs{peak})-\ln(\wobs{50}-\wobs{peak})}, \\
  k_3 & = & \frac{\wobs{peak}}{2}, \\
  k_4 & = & \left\{
  \begin{array}{ll}
       \frac{1}{4}\,\frac{\wobs{peak}^2\,{\Psi^{\rm obs}_{\rm max}}^2}{{\Psi^{\rm obs}_{\rm max}}^2-{\Psi^{\rm obs}_0}^2}\quad                          & {\rm if~}\Psi^{\rm obs}_{\rm max}>\Psi^{\rm obs}_0, \\
       0 & {\rm if~}\Psi^{\rm obs}_{\rm max}=\Psi^{\rm obs}_0,
  \end{array} \right. \\
  k_5 & = & \Psi^{\rm obs}_0\,\sqrt{k_4}.\label{eqk5}
\eea

Fig.~\ref{fig_recover_profile} compares a simulated normalized \ha-emission line with the emission line recovered using Eqs.~(\ref{eqpsiapprox}--\ref{eqk5}).

We emphasize that the approximation of Eq.~(\ref{eqpsiapprox}) does not conserve the normalization condition $\int_{\infty}^{\infty}\Psi(V)=1$. However, the integral $\int_{\infty}^{\infty}\Psi_{\rm approx}(V)$ normally differs from unity by no more than 10\%. Hence, a good approximation of the flux profile can be obtained by multiplying $\Psi_{\rm approx}(V)$ with the velocity-integrated line flux of the respective emission line as given in the simulation catalog.

We note that some emission lines, especially those of galaxies seen face-on, peak at the line center. These lines have $\wobs{peak}=0$, and therefore Eq.~(\ref{eqpsiapprox}) reduces to the exponential tails.

\newpage
\section{B. Online-access to the sky simulation}\label{appendix_access}

One particular realization of the mock observing cone, i.e.~one choice of random symmetry operations for the replicated simulation boxes (see Section \ref{subsection_mockcone}), can be access on-line via http://s-cubed.physics.ox.ac.uk/ (go to ``S$^3$-SAX-Sky''). Each galaxy in the virtual observing cone is specified by a list of properties, including its position, its attributes for the \ha- and CO-emission lines, as well as its intrinsic properties of the DeLucia-catalog, such as optical magnitudes, masses, star formation rates, clustering properties, or merger histories.

The accessible database contains two subsets, a full sky simulation ($\sim2.8\cdot10^8$ galaxies), associated with the Millennium Simulation ($\simbox=500\,\h^{-1}\,{\rm Mpc}$), and a small sky simulation ($\sim4.5\cdot10^6$ galaxies), associated with the Milli-Millennium Simulation ($\simbox=62.5\,\h^{-1}\,{\rm Mpc}$). The solid angle subtended by the small sky simulation is 64-times smaller than of the full sky simulation. However, the small simulation can be useful for testing purposes, since it can be accessed and post-processed 10--100 times faster.

The maximal opening angle $\varphi$ of the sky field depends on the maximal comoving distance $\dcmax$ (or the maximal redshift $\zmax$) via Eq.~\ref{eqphiboxes} (see also Fig.~\ref{fig_fov}). The user must be aware that there are no galaxies outside this maximal opening angle, i.e.~galaxies only exist, where ${\rm \abs{RA}}$ and ${\rm \abs{Dec}}$ are smaller than $\varphi/2$.

The database can be queried using the structured query language (SQL) interface. The latter not only allows the user to download a particular galaxy sample, but it also offers ways to directly calculate $dN/dz$-functions and luminosity-functions or to retrieve all the galaxies of a particular cluster. Samples of such advanced queries are given on the web-page.

\newpage

\section{C. Illustration of a sky field in different redshift bins}\label{appendix_illstration}

\begin{figure*}[h]
  \includegraphics[width=\textwidth]{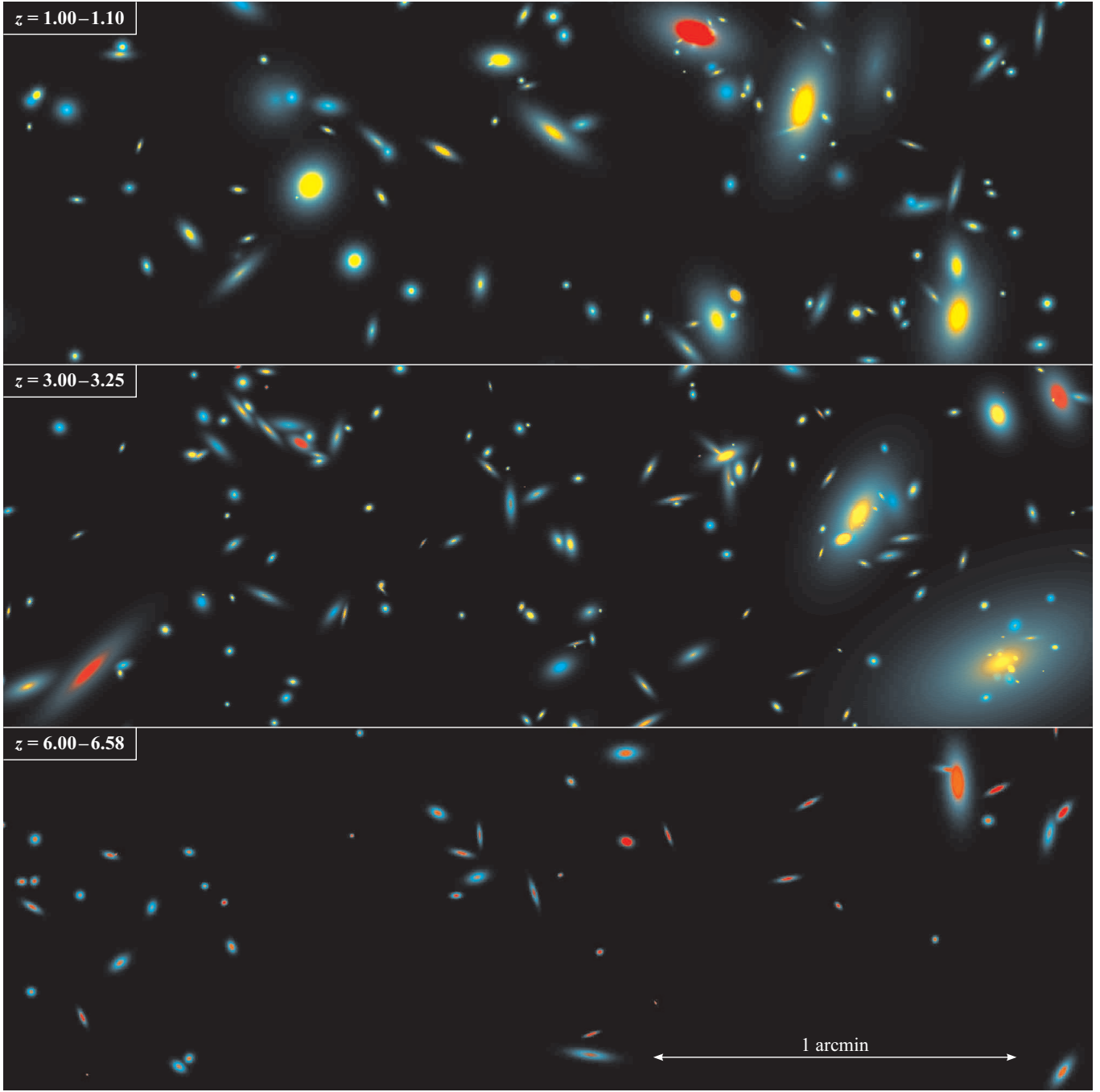}
  \caption{Simulated sky field covering $3\times1\rm~arcmin^2$. The three panels correspond to three different redshift slices, with an identical comoving depth of $240\rm~Mpc$. The coloring is identical to Fig.~\ref{fig_z1}, but the flux scales are 10-times smaller at $z=3$ and 100-times smaller at $z=6$.}\label{fig_bigmap}
\end{figure*}

Fig.~\ref{fig_bigmap} shows the HI and CO of the galaxies in a mock sky field of $3\times1\rm~arcmin^2$ at three different redshifts. Each redshift slice has the same comoving thickness of $240\rm\,Mpc$, such that the number of galaxies is proportional to the comoving space density of galaxies. Note, however, that the flux scales differ between the three panels of Fig.~\ref{fig_bigmap}. The main features of the galaxies in Fig.~\ref{fig_bigmap} are discussed in Section \ref{subsection_graphical_overview}.

\newpage

\section{D. Analytic fits for $\dndz$-functions}\label{appendix_parameters}

Table \ref{tab_parameters} lists the values of the parameters $c_1$, $c_2$, and $c_3$ for the analytic $\dndz$-fit of Eq.~(\ref{eqdndzfit}). The parameters are given for both peak flux density limited and integrated flux limited surveys, although the functions have only been displayed for the peak flux density limited case. Since apparent line widths are typically of order $100\rm\,km\,s^{-1}$, surveys limited by a flux density $s_{\rm lim}$ and those limited by a velocity-integrated flux $\svel_{\rm lim}=s_{\rm lim}\times100\rm\,km\,s^{-1}$ have similar $\dndz$-functions.

The sensitivity of a survey may be high enough that galaxies are detected in regions of the galaxy MF, where the simulation is incomplete. In fact, for each emission line and each sensitivity limit, there is a critical redshift $\zc$, below which sources of the incomplete part of the simulated galaxy MF will be detected at a sufficiently high rate that the number of real detections will significantly exceed the number of simulated detections. Therefore, the simulated $\dndz$-functions at $z\leq\zc$ must be considered as lower limits. As in Section \ref{subsection_dndz}, we shall define $\zc$ as the redshift, below which the fraction of sources with hydrogen masses below $10^8\,\msun$ (completeness limit) is larger than $1\%$, noting that the fraction of actually missing sources may be much larger due to incompleteness (see Section \ref{subsection_dndz}).

\begin{table}[h]
\centering
 \begin{tabular}{c|c|cccccc|cccccc}
 \hline
 \multicolumn{2}{c|}{\multirow{2}{*}{Parameters}} & \multicolumn{6}{c|}{\textbf{Limiting peak flux density $\mathbf{\rm [Jy]}$}} & \multicolumn{6}{c}{\textbf{Limiting integrated flux $\mathbf{\rm [Jy\,km\,s^{-1}]}$}} \\
 \multicolumn{2}{c|}{}                            & $10^{-8}$ & $10^{-7}$ & $10^{-6}$ & $10^{-5}$ & $10^{-4}$ & $10^{-3}$ & $10^{-6}$ & $10^{-5}$ & $10^{-4}$ & $10^{-3}$ & $10^{-2}$ & $10^{-1}$ \\
  \hline
            & $c_1$ &  6.55 &  6.87 &  6.73 &  5.75 &  4.56 &  6.62 &  6.61 &  6.94 &  6.53 &  5.79 &  5.21 &  6.33 \\
 \ha        & $c_2$ &  2.54 &  2.85 &  2.32 &  1.14 &  0.43 &  2.64 &  2.66 &  2.91 &  2.09 &  1.33 &  1.23 &  2.60 \\
            & $c_3$ &  1.42 &  2.17 &  3.09 &  3.95 &  6.86 & 35.49 &  1.51 &  2.34 &  2.86 &  3.92 &  8.38 & 30.52 \\
            & $\zc$ &   2.3 &   1.1 &   0.5 &   0.2 &   0.1 &   0.1 &   2.1 &   0.9 &   0.4 &   0.1 &   0.1 &   0.1 \\
 \hline
            & $c_1$ &  6.32 &  6.40 &  6.37 &  5.92 &  5.06 &  4.15 &  6.31 &  6.37 &  6.30 &  5.91 &  5.27 &  5.17 \\
 CO(1--0)   & $c_2$ &  2.16 &  2.29 &  2.05 &  1.33 &  0.54 &  0.31 &  2.14 &  2.23 &  1.96 &  1.43 &  0.95 &  1.37 \\
            & $c_3$ &  1.11 &  1.27 &  1.46 &  1.55 &  1.72 &  3.46 &  1.11 &  1.23 &  1.38 &  1.47 &  1.69 &  5.19 \\
            & $\zc$ &   6.0 &   5.1 &   2.2 &   0.9 &   0.3 &   0.1 &   6.0 &   5.1 &   2.2 &   0.9 &   0.4 &   0.1 \\
 \hline
            & $c_1$ &  6.29 &  6.34 &  6.42 &  6.25 &  5.64 &  4.64 &  6.29 &  6.33 &  6.38 &  6.21 &  5.67 &  4.87 \\
 CO(2--1)   & $c_2$ &  2.10 &  2.20 &  2.27 &  1.81 &  1.07 &  0.35 &  2.09 &  2.18 &  2.18 &  1.84 &  1.23 &  0.70 \\
            & $c_3$ &  1.07 &  1.15 &  1.34 &  1.51 &  1.58 &  1.94 &  1.06 &  1.14 &  1.28 &  1.46 &  1.49 &  1.82 \\
            & $\zc$ &   5.8 &   6.0 &   3.8 &   1.5 &   0.6 &   0.2 &   5.8 &   6.0 &   3.8 &   1.5 &   0.6 &   0.2 \\
 \hline
            & $c_1$ &  6.28 &  6.32 &  6.40 &  6.32 &  5.79 &  4.86 &  6.28 &  6.31 &  6.37 &  6.25 &  5.81 &  5.09 \\
 CO(3--2)   & $c_2$ &  2.08 &  2.16 &  2.27 &  1.95 &  1.22 &  0.47 &  2.08 &  2.14 &  2.21 &  1.89 &  1.36 &  0.87 \\
            & $c_3$ &  1.06 &  1.12 &  1.27 &  1.44 &  1.48 &  1.58 &  1.06 &  1.11 &  1.23 &  1.36 &  1.43 &  1.56 \\
            & $\zc$ &   5.8 &   6.0 &   4.9 &   2.0 &   0.7 &   0.3 &   5.8 &   6.0 &   4.7 &   2.0 &   0.8 &   0.3 \\
 \hline
            & $c_1$ &  6.28 &  6.31 &  6.37 &  6.26 &  5.74 &  4.84 &  6.28 &  6.30 &  6.34 &  6.19 &  5.77 &  5.03 \\
 CO(4--3)   & $c_2$ &  2.08 &  2.13 &  2.22 &  1.87 &  1.18 &  0.52 &  2.08 &  2.12 &  2.17 &  1.79 &  1.34 &  0.86 \\
            & $c_3$ &  1.06 &  1.10 &  1.21 &  1.32 &  1.36 &  1.38 &  1.06 &  1.09 &  1.18 &  1.25 &  1.33 &  1.36 \\
            & $\zc$ &   5.8 &   6.0 &   5.2 &   2.1 &   0.7 &   0.3 &   5.8 &   6.0 &   5.2 &   2.0 &   0.8 &   0.3 \\
 \hline
            & $c_1$ &  6.28 &  6.30 &  6.32 &  6.05 &  5.45 &  4.53 &  6.28 &  6.29 &  6.30 &  6.00 &  5.53 &  4.74 \\
 CO(5--4)   & $c_2$ &  2.07 &  2.11 &  2.13 &  1.56 &  0.96 &  0.54 &  2.07 &  2.11 &  2.09 &  1.52 &  1.20 &  0.83 \\
            & $c_3$ &  1.05 &  1.09 &  1.17 &  1.15 &  1.14 &  1.14 &  1.05 &  1.08 &  1.15 &  1.09 &  1.17 &  1.15 \\
            & $\zc$ &   5.8 &   6.0 &   5.1 &   1.7 &   0.6 &   0.2 &   5.8 &   6.0 &   4.7 &   1.6 &   0.6 &   0.2 \\
 \hline
            & $c_1$ &  6.28 &  6.29 &  6.17 &  5.62 &  4.95 &  4.17 &  6.28 &  6.28 &  6.13 &  5.62 &  5.08 &  4.35 \\
 CO(6--5)   & $c_2$ &  2.07 &  2.09 &  1.81 &  1.04 &  0.78 &  0.93 &  2.07 &  2.07 &  1.76 &  1.11 &  1.02 &  1.13 \\
            & $c_3$ &  1.05 &  1.09 &  1.07 &  0.89 &  0.92 &  1.08 &  1.05 &  1.08 &  1.03 &  0.87 &  0.97 &  1.06 \\
            & $\zc$ &   5.8 &   6.0 &   6.2 &   0.9 &   0.3 &   0.1 &   5.8 &   6.0 &   6.0 &   0.9 &   0.4 &   0.1 \\
 \hline
            & $c_1$ &  6.27 &  6.22 &  5.72 &  5.01 &  4.45 &  3.91 &  6.27 &  6.20 &  5.70 &  5.08 &  4.55 &  4.06 \\
 CO(7--6)   & $c_2$ &  2.06 &  1.96 &  1.14 &  0.69 &  1.13 &  1.58 &  2.05 &  1.93 &  1.18 &  0.80 &  1.18 &  1.72 \\
            & $c_3$ &  1.05 &  1.06 &  0.80 &  0.65 &  0.88 &  1.24 &  1.05 &  1.04 &  0.79 &  0.66 &  0.86 &  1.14 \\
            & $\zc$ &   6.0 &   5.2 &   1.3 &   0.4 &   0.2 &   0.1 &   6.0 &   5.2 &   1.2 &   0.4 &   0.1 &   0.1 \\
 \hline
            & $c_1$ &  6.22 &  5.76 &  5.00 &  4.42 &  4.13 &  3.68 &  6.20 &  5.74 &  5.08 &  4.48 &  4.19 &  3.78 \\
 CO(8--7)   & $c_2$ &  1.97 &  1.25 &  0.66 &  1.04 &  1.85 &  2.22 &  1.95 &  1.27 &  0.80 &  0.99 &  1.83 &  2.06 \\
            & $c_3$ &  1.04 &  0.78 &  0.51 &  0.62 &  1.03 &  1.45 &  1.02 &  0.78 &  0.55 &  0.58 &  0.97 &  1.19 \\
            & $\zc$ &   6.0 &   6.6 &   0.5 &   0.2 &   0.1 &   0.1 &   6.0 &   1.3 &   0.5 &   0.2 &   0.1 &   0.1 \\
 \hline
            & $c_1$ &  5.70 &  4.95 &  4.35 &  4.10 &  3.84 &  3.33 &  5.69 &  5.03 &  4.43 &  4.12 &  3.93 &  3.43 \\
 CO(9--8)   & $c_2$ &  1.24 &  0.74 &  1.05 &  1.81 &  2.31 &  2.17 &  1.27 &  0.87 &  1.04 &  1.70 &  2.20 &  2.27 \\
            & $c_3$ &  0.73 &  0.46 &  0.48 &  0.81 &  1.14 &  1.43 &  0.73 &  0.51 &  0.48 &  0.73 &  1.04 &  1.21 \\
            & $\zc$ &   6.2 &   0.4 &   0.1 &   0.1 &   0.1 &   0.1 &   1.2 &   0.4 &   0.1 &   0.1 &   0.1 &   0.1 \\
 \hline
            & $c_1$ &  4.81 &  4.29 &  4.03 &  3.88 &  3.52 &  2.95 &  4.89 &  4.35 &  4.07 &  3.91 &  3.65 &  3.00 \\
 CO(10--9)  & $c_2$ &  0.86 &  1.29 &  1.80 &  2.29 &  2.27 &  2.23 &  0.95 &  1.26 &  1.76 &  2.21 &  2.35 &  2.03 \\
            & $c_3$ &  0.42 &  0.48 &  0.66 &  0.97 &  1.11 &  1.47 &  0.47 &  0.48 &  0.64 &  0.89 &  1.06 &  1.06 \\
            & $\zc$ &   0.3 &   0.1 &   0.1 &   0.1 &   0.1 &   0.1 &   0.3 &   0.1 &   0.1 &   0.1 &   0.1 &   0.1 \\
 \hline
 \end{tabular}
\caption{Parameters for the analytic fit formula of Eq.~(\ref{eqdndzfit}) for $\dndz$ peak flux density limited and integrated flux limited surveys.}
\label{tab_parameters}
\end{table}

\end{document}